\documentclass{emulateapj}

\def\degpnt{^{\circ}\kern-1.7mm.\kern+.35mm}
\def\arcpnt{"\kern-1.7mm.\kern+.35mm}
\def\minpnt{'\kern-1.0mm.\kern+.30mm}

\slugcomment{Accepted to ApJ}

\shorttitle{SNe~Ia in Local Hosts}
\shortauthors{Neill et al.}

\received{\today}
\begin{document}


\title{The Local Hosts of Type Ia Supernovae}


\author{James~D.~Neill\altaffilmark{1}, Mark~Sullivan\altaffilmark{2}
	D.~Andy~Howell\altaffilmark{3}, Alex~Conley\altaffilmark{4},
	Mark~Seibert\altaffilmark{5}, D.~Christopher Martin\altaffilmark{1},
	Tom A. Barlow\altaffilmark{1}, Karl Foster\altaffilmark{1},
	Peter G. Friedman\altaffilmark{1}, Patrick Morrissey\altaffilmark{1},
	Susan G. Neff\altaffilmark{6}, David Schiminovich\altaffilmark{7},
	Ted K. Wyder\altaffilmark{1}, 
	Luciana  Bianchi\altaffilmark{8}, Jos\'e Donas\altaffilmark{9},
	Timothy M. Heckman\altaffilmark{10}, Young-Wook Lee\altaffilmark{11},
	Barry F. Madore\altaffilmark{5}, Bruno Milliard\altaffilmark{9},
	R. Michael Rich\altaffilmark{12}, and Alex S. Szalay\altaffilmark{10}
}

\altaffiltext{1}{California Institute of Technology, 1200 E. California Blvd., Pasadena, CA 91125, USA}
\altaffiltext{2}{University of Oxford, Denys Wilkinson Building,
Keble Road, Oxford, OX1 3RH, UK}
\altaffiltext{3}{Las Cumbres Observatory Global Telescope Network, 6740 Cortona Dr., Suite 102, Goleta, CA 93117, USA}
\altaffiltext{4}{Department of Astronomy and Astrophysics, University of 
Toronto, 50 St. George Street, Toronto, ONM5S3H8, Canada}
\altaffiltext{5}{The Observatories of the Carnegie Institute of Washington, 813 Santa Barbara Street, Pasadena, CA, 91101, USA}
\altaffiltext{6}{Laboratory for Astronomy and Solar Physics, NASA   
Goddard Space Flight Center, Greenbelt, MD, 20771}
\altaffiltext{7}{Department of Astronomy, Columbia University, New   
York, NY 10027}
\altaffiltext{8}{Center for Astrophysical Sciences, The Johns Hopkins  
University, 3400 N. Charles St., Baltimore, MD, 21218}
\altaffiltext{9}{Laboratoire d'Astrophysique de Marseille, BP 8,  
Traverse du Siphon, 13376 Marseille Cedex 12, France}
\altaffiltext{10}{Department of Physics and Astronomy, The Johns  
Hopkins University, Homewood Campus, Baltimore, MD 21218}
\altaffiltext{11}{Center for Space Astrophysics, Yonsei University,  
Seoul 120-749, Korea}
\altaffiltext{12}{Department of Physics and Astronomy, University of  
California, Los Angeles, CA 90095}


\begin{abstract}

We use multi-wavelength, matched aperture, integrated photometry from {\it
GALEX}, SDSS and the RC3 to estimate the physical properties of 166 nearby
galaxies hosting 168 well-observed Type Ia supernovae (SNe~Ia).  The
ultra-violet (UV) imaging of local SN~Ia hosts from {\it GALEX} allows a
direct comparison with higher redshift hosts measured at optical
wavelengths that correspond to the rest-frame UV.  Our data corroborate
well-known features that have been seen in other SN~Ia samples.
Specifically, hosts with active star formation produce brighter and slower
SNe~Ia on average, and hosts with luminosity-weighted ages older than 1 Gyr
produce on average more faint, fast and fewer bright, slow SNe~Ia than
younger hosts.  New results include that in our sample, the faintest and
fastest SNe~Ia occur only in galaxies exceeding a stellar mass threshhold
of $\sim10^{10}$ M$_{\odot}$, leading us to conclude that their progenitors
must arise in populations that are older and/or more metal rich than the
general SN~Ia population.  A low host extinction sub-sample hints at a
residual trend in peak luminosity with host age, after correcting for
light-curve shape, giving the appearance that older hosts produce
less-extincted SNe~Ia on average.  This has implications for cosmological
fitting of SNe~Ia and suggests that host age could be useful as a parameter
in the fitting.  Converting host mass to metallicity and computing
$^{56}$Ni mass from the supernova light curves, we find that our local
sample is consistent with a model that predicts a shallow trend between
stellar metallicity and the $^{56}$Ni mass that powers the explosion, but
we cannot rule out the absence of a trend.  We measure a correlation
between $^{56}$Ni mass and host age in the local universe that is shallower
and not as significant as that seen at higher redshifts.  The details of
the age -- $^{56}$Ni mass correlations at low and higher redshift imply a
luminosity-weighted age threshhold of $\sim3$ Gyr for SN~Ia hosts, above
which they are less likely to produce SNe~Ia with $^{56}$Ni masses above
$\sim0.5$ M$_{\odot}$.

\end{abstract}




\section{Introduction\label{sec_intro}}

A fundamental motivation for studying Type Ia supernovae (SNe~Ia) is to arrive
at a physical explanation for the observed fact that these objects have
well-behaved and calibratable explosions \citep{Phillips:93:L105}.  This
property gives cosmographers a tool for measuring universal expansion and
provided the first direct observational evidence for cosmic acceleration
\citep{Riess:98:1009, Perlmutter:99:565}.  Knowing the physical process that
generates SNe~Ia provides strong constraints on the nature of the SN~Ia
progenitor.  This, in turn, allows us to predict the evolution of the SN~Ia
population \citep[e.g.,][]{Howell:07:L37}.  This results in tighter control of
the systematic errors arising from population evolution with redshift, an
important uncertainty in measuring cosmological parameters
\citep[e.g.,][]{Astier:06:31,Conley:09}.  A well-constrained physical model for
SNe~Ia would have further utility in tracing high-redshift star formation and
for predicting the effects of SNe~Ia on the chemical enrichment of their host
galaxies \citep[e.g.,][]{Kobayashi:07:215}.

In order to calibrate SNe~Ia and make them useful for cosmology, their
intrinsic variation must be measured and accounted for.  The peak absolute
magnitude of a given SN~Ia is a strong function of its initial decline rate
\citep{Phillips:93:L105}, and a weaker function of its peak color
\citep{Riess:96:88, Tripp:98:815, Tripp:99:209, Parodi:00:634}.  These
empirical correlations reduce the intrinsic variation of $\sim$1 magnitude
in the B-band to $\sim0.1$ magnitudes and thus provide the accurate
luminosity distance estimates required for measuring cosmological
parameters \citep[e.g.,][]{Riess:96:88, Tonry:03:1, Guy:05:781,
Prieto:06:501, Jha:07:122, Conley:08:482}.  The goal of SN~Ia progenitor
studies is to use these measures of SN~Ia intrinsic variation to explore
trends with host properties that could shed light on the underlying
population that produces SNe~Ia \citep{Phillips:93:L105, Hamuy:95:1,
Hamuy:00:1479, Howell:01:L193, Gallagher:05:210, Sullivan:06:868,
Gallagher:08:752}.

Since SN~Ia explosions are powered by the radioactive decay of $^{56}$Ni,
it is widely accepted that the intrinsic variation in SN~Ia brightness and
decline rate is primarily driven by the amount of $^{56}$Ni present in the
SN explosion, with more luminous and slower declining explosions being
powered by more $^{56}$Ni \citep{Truran:67:2315, Colgate:69:623}.  One
avenue of exploration would be to connect this theoretical idea to the
progenitor population of SNe~Ia by comparing host galaxy properties with
the measures of SN~Ia light curve variation.  This requires a mechanism
that varies the amount of $^{56}$Ni as a function of some property of the
host galaxy.

\citet{Timmes:03:L83} presented a model relating progenitor metallicity to
produced SN~Ia $^{56}$Ni mass for a constant (Solar) O/Fe ratio.  This
model was expanded and tested by \citet[hereafter H09]{Howell:09:661} using
an intermediate redshift ($0.2 < z < 0.75$) sample of SNe~Ia.  They used
spectral energy distribution (SED) fitting of optical integrated host
photometry to infer the host stellar masses for their sample.  These masses
were used as a proxy for host metallicities through the
\citet{Tremonti:04:898} mass-metallicity relation.  They estimated
the peak bolometric luminosity and rise time from the SN~Ia photometry to
calculate the required mass of $^{56}$Ni to power the explosion and
compared the calculated host metallicity versus Ni mass trend to the
\citet{Timmes:03:L83} model.  Their data are consistent with the model,
although with a higher scatter than predicted by the theory.  They also
found that varying the O/Fe ratio according to thin or thick-disk models
produced no substantial change in their results.

If the dependance of $^{56}$Ni mass on host metallicity is real, then
fainter and faster declining SNe~Ia should be associated with higher
metallicity, and thus more massive, host galaxies.  If a decline in the
rate of fast fading, fainter SNe~Ia with redshift could be detected, it
would be evidence for the relation between Ni mass and metallicity because
we expect the more distant universe to consist of lower-mass and
lower-metallicity objects on average when compared with the local universe.
At this time, however, Malmquist biases complicate an accurate
determination of the rate evolution for low-luminosity SNe~Ia
\citep{Foley:08,Gaitan:09}.  We therefore look to host properties of
low-luminosity SNe~Ia in the local universe, where they are easier to
detect and apparently more abundant.  We use this larger sample to see if
the fainter SNe do appear in higher mass, higher metallicity galaxies on
average.

To compare host properties with SN~Ia variation, we construct a
low-redshift sample of SNe~Ia with well observed light curves that provide
a light curve shape parameter, called stretch, the observed maximum light
color, and the peak luminosity of the SN in the rest-frame B-band
\citep{Conley:08:482}.  We characterize the hosts of these SNe using newly
generated and catalog galaxy integrated UV and optical magnitudes measured
within matched apertures as inputs to a SED fitting program that estimates
host stellar mass, luminosity-weighted age and star formation rate (SFR).
We first use these results to explore the relation between the observed
light curve parameters and the derived host properties and compare our
results to studies at higher redshift using the same methods
\citep[][H09]{Sullivan:06:868}.  We isolate a subset of hosts with low
internal extinction in order to explore the relationship between SN peak
brightness and host properties where the source of line-of-sight color for
the SNe is minimized.  We then extend the work of H09 by applying their
techniques to our local sample to derive SN~Ia $^{56}$Ni-mass, from the
light-curve photometry, and compare these with host metallicities, from our
derived host stellar masses, and host luminosity-weighted ages.  Throughout
this paper we express host masses in M$_{\odot}$ and SFR in M$_{\odot}$
yr$^{-1}$ and assume a Hubble constant of $H_0 = 73$ km s$^{-1}$
Mpc$^{-1}$.

\section{Sample Selection and Analysis}

Here we outline the properties of our low-redshift SN~Ia sample and
describe the methods we use to characterize the SN and host galaxy
properties.

\subsection{The Local SN Sample and Light Curve Fitting\label{sec_sample}}

For our local sample we start with the low-z literature sample summarized
in \citet{Conley:08:482}.  We supplement this with more recent SN
photometry from \citet{Hicken:09b}.  The initial pool of nearby SNe~Ia are
required to have modern CCD photometry with errorbars, are required to have
phase coverage beyond 3 days after maximum light and prior to 7 days after
maximum light.  They are also required to have one restframe B-band
magnitude between 8 days before and 13 days after maximum light and one
restframe U or V-band magnitude in the same phase range.  In addition the
wavelength range of the filters must be between 2700 and 7200 \AA.

To be consistent with H09 we adopt the SN~Ia light curve fitting method
from \citet{Conley:08:482}.  This method fits the SN light curve data to
produce the stretch, $s$ \citep{Perlmutter:97:565}, a peak restframe B-band
apparent magnitude, m$_{Bmax}$, and a SN color, $\mathcal{C}$, which is
somewhat like $(B-V)_{Bmax}$ in the sense that redder SNe have higher
$\mathcal{C}$ values.  Briefly, the stretch parameter is a time-axis scale
factor that is applied to the light curve that aligns its shape with a
`canonical' SN~Ia light curve.  The sense of the stretch parameter is such
that faster, fainter SNe~Ia have low stretch values and brighter, slower
SNe~Ia have higher stretch values.  The sense of the SN color parameter is
such that bluer SNe at maximum are brighter.  The corrected peak brightness
of the SN~Ia used for cosmology is determined by a linear combination of
$s$, $\mathcal{C}$, and m$_{Bmax}$ \citep{Astier:06:31, Conley:08:482}.
\citet{Conley:08:482} present a much more detailed description of this
method including comparisons with other popular fitting routines.  These
comparisons demonstrate that, at least for this study, our results are not
dependant on the fitting technique.

\begin{figure}[h]
\includegraphics[scale=0.35,angle=90.]{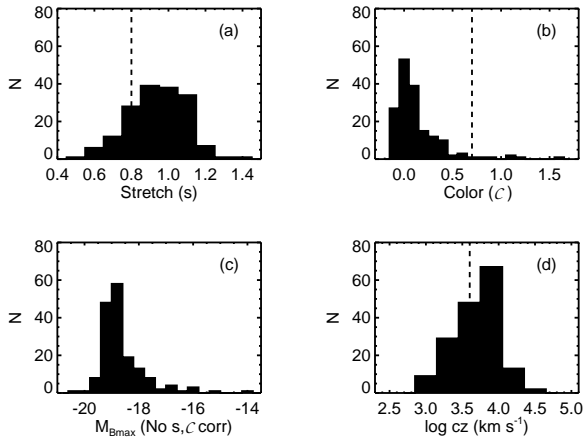}
\caption{Frequency diagrams of light curve fit parameters for the local SNe~Ia
having enough observations to determine a stretch value and having
sufficient host photometry to allow good fits to their SED: (a) stretch, (b)
color, (c) M$_{Bmax}$ (uncorreced for stretch-color), (d) recession
velocity in km~s$^{-1}$.  The
vertical dashed lines show various limits described in the text: (a) the
limit between ordinary and low-luminosity SNe~Ia ($s=0.80$), (b) the peak
color limit above which the fitting becomes less accurate
($\mathcal{C}=0.7$), and (d) the Hubble flow limit adopted here ($cz =
4000$~km~s$^{-1}$, $z = 0.013$).  There are 168 SNe~Ia in the sample with 
well-fit host SEDs.
\label{fig_smpldist}
}
\end{figure}

Figure~\ref{fig_smpldist} shows distributions of the fitted SN light curve
properties for all the low-redshift SNe passing the previously described
criteria in hosts with sufficient integrated host photometry to provide
good SED fits (168 SNe~Ia, filled histograms).  This is the base sample we
will examine in this work.  The figure shows the distributions in stretch,
peak color, absolute B magnitude (uncorrected for stretch-color), and
recession velocity.  The light curve fit properties of this sample are
given in Table~\ref{tab_sne} which lists the SN name, the host galaxy, the
recession velocity, the light curve stretch, the fitted apparent B
magnitude at maximum light, and a status column which is described below.
More details of the light-curve fits for these SNe can be found in
\citet{Conley:09}.

Low-luminosity SNe~Ia are generally better represented in lower redshift
samples for multiple reasons.  The spectra and the stretch-luminosity and
color-luminosity relations for low stretch SNe~Ia differ enough from
higher-stretch SNe \citep{Garnavich:04:1120,Taubenberger:08:75} to increase
the errors in fitting their light curves beyond the error-budget
requirement for cosmological parameter determination.  This, combined with
their faintness, makes low-stretch SNe less appealing targets for the
spectroscopic followup required for cosmological surveys.  As a result, the
SN~Ia training set derived for cosmology in \citet{Conley:08:482} contains
no SNe with $s < 0.70$, and very few below $s = 0.75$.  At lower redshifts
where even faint SNe are relatively easy to followup, the unusual objects
tend to attract attention and acquire spectroscopy.

The break in the stretch-luminosity relation appears to be at $s\sim0.80$
\citep{Gaitan:09}.  For this paper, we define the set of low-luminosity
SNe~Ia based on stretch to be those with $s < 0.80$ (to the left of the
dashed vertical line in Figure~\ref{fig_smpldist}a).  The training set from
\citet{Conley:08:482} allows us to extend this limit slightly to $s = 0.75$
when we apply corrections for stretch-luminosity.

Our local sample contains 29 SNe with $s < 0.80$ of which 12 are below $s =
0.70$.  Since we are using a light curve method developed for cosmology,
this means our fitting can only tell that an object with $s < 0.70$ is
low-stretch, but cannot accurately measure the stretch.  We have increased
the error bars by a factor of three for objects with $s < 0.70$.  The
sample used in H09 was derived from the Supernova Legacy Survey
\citep[SNLS,][]{Astier:06:31} which was optimized for cosmology.  Either
due to the fact that such objects are very rare beyond $z = 0.2$, or
because the spectroscopic followup was biased against them or both, there
are no objects with $s < 0.75$ in the H09 sample \citep[see
also][]{Bronder:08:717}.  The low stretch SNe~Ia in our sample are
indicated with an `L' in the status column in Table~\ref{tab_sne}.

A similar situation exists for the peak color, $\mathcal{C}$.  SNe~Ia with
colors greater than $\mathcal{C} \sim 0.7$ (see
Figure~\ref{fig_smpldist}b) are also unappealing objects for cosmological
surveys because they are fainter and because they have larger fitting
errors.  These red SNe~Ia should not be excluded when examining host
properties, however.  Our sample contains six SNe~Ia with $\mathcal{C}>0.7$.
These SNe are indicated with an `R' in the status column of
Table~\ref{tab_sne}.  We have increased the peak color errors by a factor
of ten for these SNe to reflect the higher uncertainty in this parameter.

We emphasize that while the local sample assembled here is appropriate for
exploring the link between host and SN properties, many of the SNe in our
sample are too local ($cz<4000$~km~s$^{-1}$, $z < 0.013$, to the left of
the vertical dashed line in Figure~\ref{fig_smpldist}d) or have stretch
errors or colors large enough to exclude them from use in cosmological
fits.  We indicate the cosmology status of our sample SNe~Ia in the last
column of Table~\ref{tab_sne} where an H indicates a SN~Ia in the Hubble
flow ($cz>4000$~km~s$^{-1}$), and a C indicates a SN~Ia that is eligible
for cosmological fitting based on the light-curve fit quality and having
parameters within the range where the light curve fitting has reasonably
low errors.

\subsection{Integrated Host Photometry and SED fitting\label{sec_host_phot}}

We assembled host integrated photometry in the ultraviolet (UV) and optical
to characterize the SN~Ia host SEDs.  For our low-redshift host galaxy
sample, the addition of the {\it GALEX} UV photometry improves the
estimation of short-term star formation
\citep[e.g.,][Figure~1]{Martin:05:L1}, and provides similar constraints on
recent star formation when compared with the higher-redshift samples from
the SNLS, whose blue optical photometry corresponds to the rest-frame UV.
All host magnitudes are total magnitudes using matched elliptical apertures
having the standard D25 diameter, to be compatable with the RC3.  All host
magnitudes are corrected for Milky Way (foreground) extinction using the
dust maps of \citet{Schlegel:98:525}.

In the UV, we generated host integrated magnitudes as part of the
preliminary efforts to produce the {\it GALEX} Large Galaxy Atlas
\citep[GLGA,][]{Seibert:09}.  This atlas will measure $\sim$20,000 galaxies
imaged by the {\it GALEX} UV-imaging satellite \citep{Martin:05:L1} having
diameters in the UV greater than one arcminute.  The {\it GALEX} imaging
mode has two bandpasses, one in the far-UV (FUV, $\lambda_{eff} = 1539$~\AA,
$\Delta\lambda = 442$~\AA) and another in the near-UV (NUV, $\lambda_{eff}
= 2316$~\AA, $\Delta\lambda = 1060$~\AA).  We use a technique similar
to that used to generate the Nearby Galaxy Atlas \citep{Paz:07:185} with
which we cross-checked our integrated UV magnitudes.  This method will be
described in detail in \citet{Seibert:09}, but to summarize, we perform
surface photometry in elliptical apertures on sky subtracted images that
have had foreground point sources and background galaxies masked.

For optical photometry, we used the RC3 \citep{Vaucouleurs:91} and/or
images obtained from the Sloan Digital Sky Survey \citep[SDSS\footnote{
{\tt http://www.sdss.org}},][]{York:00:1579} in the five SDSS bands: {\it
ugriz}.  Our sample required either SDSS coverage or an integrated
magnitude from the RC3.  The RC3 total integrated Johnson UBV magnitudes
were obtained directly from NED\footnote{{\tt
http://nedwww.ipac.caltech.edu/}}.  For the larger hosts in our sample, we
found the SDSS catalog data to be inaccurate for a number of reasons.  Some
of these hosts spanned multiple image strips and many were broken up into
sub-regions making the determination of a total flux problematic.  In order
to enforce consistency across wavelengths we decided to coadd and mosaic
the SDSS image data for each SN host and derive the integrated photometry
ourselves.  To achieve this, we adapted our integrated photometry methods
developed for the GLGA to SDSS image data.  The major difference between
GALEX and SDSS data is in the treatment of the sky background which is
extremely low in GALEX data.  We checked the consistency of our integrated
SDSS magnitudes by comparing them with RC3 magnitudes and found them to
agree within the errors.  All UV and optical integrated galaxy magnitudes
are presented in \citet{Seibert:09}.

\citet{Sullivan:06:868} fit optical photometry to galaxy SED models
produced with the PEGASE.2 SED galaxy spectral evolution code
\citep{Fioc:97:950, Borgne:02:446, Borgne:04:881}.  We adopted this method
to allow a direct comparison with the higher redshift host studies of
\citet{Sullivan:06:868} and H09.  As with these studies, all redshifts are
known from the hosted SNe.  This has the benefit of eliminating redshift
degeneracies in the SED fits.  The average internal extinction was allowed
to vary in the fits over the range $0.0 < E(B-V)_{HOST} < 0.7$ in
increments of 0.05 magnitudes using a \citet{Calzetti:94:582} dust model
\citep[see][]{Fioc:97:950}.  We point out the stellar mass derived from
PEGASE.2 models is an estimate of the current mass in stars and is not the
integral of the star formation history for a given galaxy
\citep{Sullivan:06:868}.  We also emphasize that our ages are strongly
influenced by the flux produced by ongoing star formation yielding a fairly
tight correlation between our derived host ages and specific star formation
rates.  Thus, we should keep in mind that our luminosity-weighted properties
might be different from those produced with mass-weighted measurements.
The details of the galaxy modeling and SED fitting method employed here can
be found in \S3 of \citet{Sullivan:06:868}.

\begin{figure}[h]
\includegraphics[scale=0.38,angle=90.]{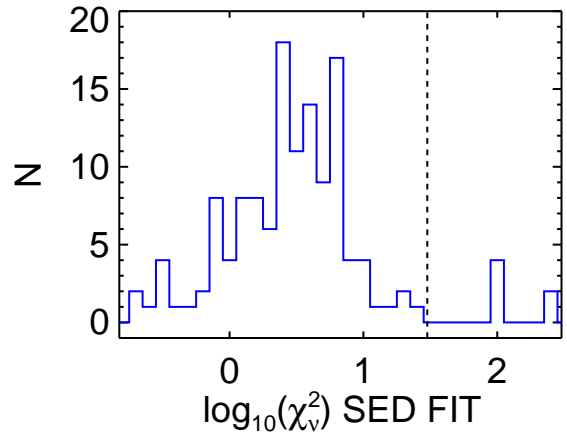}
\caption{Frequency diagram of $\chi^2_{\nu}$ of the host SED fits
illustrating one criterion for eliminating unusable hosts: we
require $\chi^2_{\nu} < 30$ for the fit.  Large values of $\chi^2_{\nu}$
arise because local hosts have smaller photometric errors than hosts at
higher redshifts.  The other criteria require a
unique solution for the SED and require UV and optical photometric points
in each SED.  These criteria reduce the sample of SNe~Ia with good light
curves from 258 down to 168 and produce 166 unique host SED fits (two
galaxies in our sample hosted two SNe~Ia each, see text).
\label{fig_sed_chi2}
}
\end{figure}

To select the sample with good SED fits, we first require that a unique
solution be found.  This reduced our initial host sample from 258 down to
209, mostly due to the rejected SED having too few photometric points.  We
next require SED coverage in both the UV and optical bands, further
reducing our sample down to 174 hosts.  At this point we examined the
distribution of $\chi^2_{\nu}$ and eliminated an additional six hosts that
had extreme values \citep[$\chi^2_{\nu} > 30$, see
Figure~\ref{fig_sed_chi2} and][]{Borgne:02:446}
leaving a total sample of 168. The 90 hosts that did not pass these
criteria had no or limited optical photometry mostly due to not being
members of any of the major galaxy catalogs (i.e., anonymous) and not being
in the SDSS footprint which would allow us to derive the integrated flux
from SDSS imaging.  A subset of these also had no UV coverage because of
proximity to a UV-bright star, which precludes {\it GALEX} observations due
to detector safety.  Table~\ref{tab_host_sed} lists the host T-type
\citep{Vaucouleurs:59:275} and the SED fit derived properties for the 166
unique hosts in our sample.

The number of unique hosts is two less than the number of SNe~Ia because
two galaxies hosted two SNe~Ia each: NGC1316 hosted SN1980N and SN1981D,
and NGC5468 hosted SN1999cp and SN2002cr.  In NGC1316, the SNe~Ia produced
are low-stretch ($s\sim8.5$) with values that differ by only 3\%, while the
two SNe~Ia produced in NGC5468 are normal stretch ($s\sim0.98$) and differ
by 11\%.  This is certainly not a conclusive test of using integrated host
magnitudes for comparison with stretch values, but it gives no cause to
doubt this method, as would be the case if these pairs of SNe~Ia differed
by greater amounts.

\begin{figure}[h]
\includegraphics[scale=0.35,angle=90.]{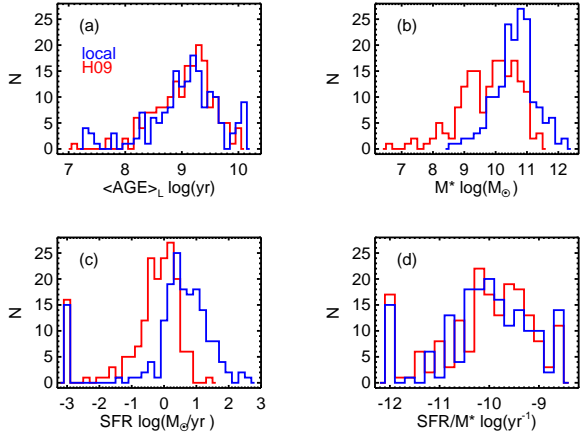}
\caption{Frequency diagrams of host properties derived from SED fitting
using the PEGASE.2 library \citep{Borgne:04:881} with the SNLS sample from
H09 as the red histograms and the local SN~Ia light-curve fit sample as the
blue histograms: (a) luminosity weighted age, (b) stellar mass, (c) star
formation rate, and (d) sSFR defined as SFR/M*.  Figure (c) illustrates our SFR
threshhold of $10^{-3}$ M$_{\odot}/$yr.  Figure (d) illustrates our sSFR
thresshold of $10^{-12}$ yr$^{-1}$.  Hosts below these values are set to
the threshhold value.  We see that the local sample is missing the
low-mass and intermediate SFR peaks seen in the H09 sample.  In sSFR, the
two distribution are quite similar.
\label{fig_pegdist}
}
\end{figure}

Figure \ref{fig_pegdist} shows the distributions of derived host properties
for the local (blue histograms) and H09 (red histograms) samples.  There
are some interesting differences between the two samples.  The discrepancy
in the mass distributions (Figure \ref{fig_pegdist}b) can be explained
because local SNe are discovered in host-targeted surveys which prefer more
massive hosts, while the SNLS is an areal survey without a mass bias and
therefore includes lower mass hosts.  The discrepancy in the SFR
distributions (Figure \ref{fig_pegdist}c) could be due to a luminosity bias
from the host-targeted low-redshift surveys.  Here the higher redshift
areal survey does not prefer luminous, and therefore higher SFR, hosts.
These differences are reinforced if there is a correlation between mass and
SFR.  In this case, the same bias that produces the deficit of low-mass
hosts in the local mass distribution produces the deficit of lower SFR
hosts in the local SFR distribution.  Figure~\ref{fig_pegdist}d shows that
when considering the specific star formation (sSFR), defined as SFR divided
by stellar mass, our distributions are fairly similar.  Thus, from here on
we consider sSFR in preference over SFR.

\section{Results}

\subsection{Host Properties versus SN Light Curve Stretch\label{sec_stretch}}

We start by examining correlations in host properties with SN~Ia light
curve stretch.  In the following discussion, it is good to keep in mind
that higher stretch SNe~Ia are brighter and lower stretch SNe~Ia are
fainter.  In these plots, we highlight the transition between the low
luminosity (i.e. low-stretch) and normal SNe with an (orange) dashed line
at $s = 0.80$.

\begin{figure}[h]
\includegraphics[scale=0.35,angle=90.]{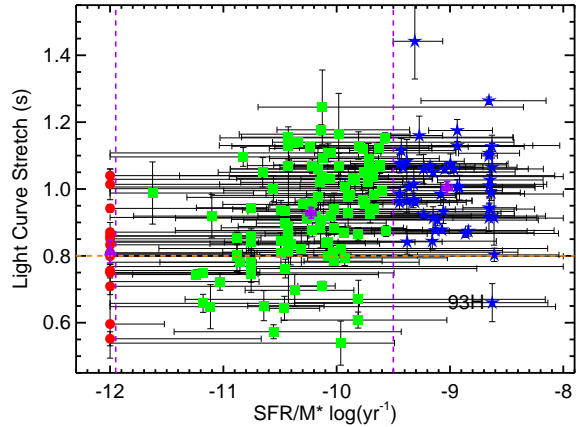}
\caption{Specific star formation rate as derived from PEGASE.2 SED fits
plotted as a function of light curve stretch for SN~Ia hosts.  Filled
circles (red) indicate hosts with very low specific star formation (sSFR 
$\leq 10^{-12}$~yr$^{-1}$), filled stars (blue) indicate hosts with strong 
specific star formation (sSFR $> 10^{-9.5}$~yr$^{-1}$) and filled squares 
(green) indicate hosts with intermediate sSFR.  Averages for the three sSFR 
bins (vertical dashed purple lines) are shown as the filled purple triangles
with the errorbars indicating the error in the mean.  The division between
normal and low-stretch SNe is indicated by the horizontal orange line at
$s=0.8$.  The one high sSFR host of a low-stretch SN~Ia (SN 1993H) is
indicated.  Stretch, on average, appears to increase with host sSFR.
\label{fig_str_ssfr}
}
\end{figure}

Figure~\ref{fig_str_ssfr} plots the log of the sSFR against hosted SN~Ia
light curve stretch.  For this plot and plots following, the symbols
indicate the host's sSFR: filled circles (red) indicate hosts with no
detected SF (sSFR $\leq 10^{-12}$~yr$^{-1}$), filled stars (blue) indicate
strongly star-forming hosts (sSFR $> 10^{-9.5}$~yr$^{-1}$) and filled
squares (green) indicate hosts with intermediate sSFR.  Typically, one
would expect elliptical galaxies to belong in the lowest sSFR group and
spiral galaxies to belong to the intermediate and high sSFR groups.  Our
aim here is to use a metric that is more physical than morphology, hence we
do not use morphological classifications to examine these SN~Ia hosts.

We notice several interesting features in this diagram.  We see that the
sensitivity threshhold of our models is just above $\sim10^{-12}$
yr$^{-1}$.  We set all hosts with no detected sSFR to this value and
consider these to be `dead' hosts.  At the high end, we see the timescale
limit which is determined by the lifetime of stars with SEDs that peak in
the UV; $\sim 3 \times 10^8$ yr.  Since all of our hosts are essentially at
zero redshift, this limit appears as a line.  The purple triangles show the
average stretch in three sSFR bins indicated by the vertical dashed lines.
Note that low stretch SNe are rare in high sSFR hosts (the one exception is
SN1993H, as indicated on the plot), but common in intermediate and low sSFR
hosts.

\begin{figure}[h]
\includegraphics[scale=0.35,angle=90.]{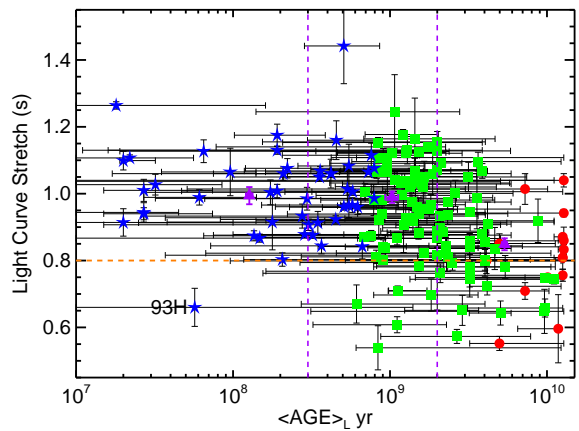}
\caption{Luminosity-weighted stellar age as derived from PEGASE.2 SED fits 
plotted as a function of light curve stretch for SN~Ia hosts.  The host
sSFR is coded as in Figure~\ref{fig_str_ssfr}.  The average in three
age bins divided at $3 \times 10^8$ and $2 \times 10^9$ yr (vertical dashed
purple lines) is shown by the filled purple triangles.  The horizontal
dashed orange line indicates the dvision between normal and low-stretch
SNe.  The one low-strech SN~Ia with a luminosity-weighted age less than $3
\times 10^8$ yr (SN 1993H) is indicated.  After a host exceeds 1 Gyr in 
age, it begins producing progressively lower stretch SNe~Ia, on average.
\label{fig_str_age}
}
\end{figure}

Figure~\ref{fig_str_age} plots the luminosity-weighted host age against
stretch, with the same symbol coding as in Figure~\ref{fig_str_ssfr}.  This
plot shows that age and sSFR are well correlated with all the high sSFR
galaxies having ages less than 1 Gyr and all the low sSFR hosts having ages
greater than 4 Gyr as expected (see \S\ref{sec_host_phot}).  The averages
(purple filled triangles) were derived from three age bins (divided at $3
\times 10^8$ and $2 \times 10^9$ yr) and show a downward trend of stretch
with age beyond 1 Gyr.  Here we find the low-stretch SNe to be evenly
distributed between host luminosity-weighted ages older than
$5\times10^8$~yr, but rare below this age with SN 1993H again being the
sole exception.

\begin{figure}[h]
\includegraphics[scale=0.35,angle=90.]{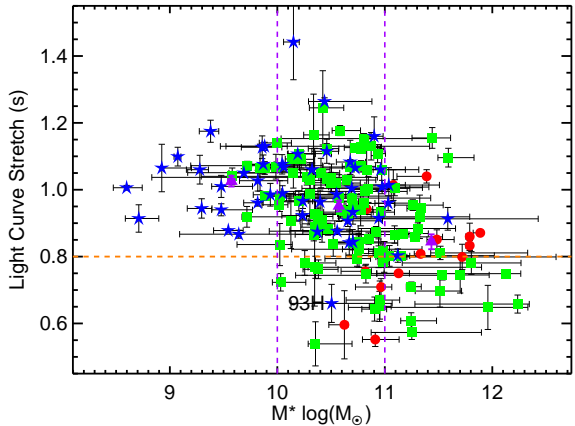}
\caption{Stellar mass as derived from PEGASE.2 SED fits plotted as a 
function of light curve stretch for SN~Ia hosts.  The host sSFR is coded as
in Figure~\ref{fig_str_ssfr}.  The averages of the sample divided at 
$\log$ M$_* = 10$ and $11$ (vertical dashed purple lines) are shown by the 
purple filled triangles.  There are no hosts of low stretch SNe with masses 
less than $10^{10}$ M$_{\odot}$.  The exceptional SN 1993H is indicated
(see text).  This figure is very similar to Figure~4 
in H09 derived from a higher redshift sample, but here we have a larger 
sample of low-stretch SNe~Ia.
\label{fig_str_mass}
}
\end{figure}

Figure~\ref{fig_str_mass} plots host stellar mass, M$_*$, against light
curve stretch and shows that hosts of the lowest stretch SNe~Ia tend to
have stellar masses higher than $10^{10}$~M$_{\odot}$.  The higher stretch
SNe~Ia hosts have a much wider mass range ($10^8$ to
$10^{12}$~M$_{\odot}$).

\subsection{SN~Ia Light Curve Color\label{sec_color}}

There is a major challenge in interpreting SN~Ia peak color because of the
unknown mix of two potential sources of the color: intrinsic SN~Ia color
from the details of the explosion itself, and host galaxy line-of-sight
reddening.  In other words, the host extinction is not accounted for in the
fitting process, which means that the peak color contains an intrinsic
component and a host extinction component.  This difficulty manifests
itself when the color-luminosity relation is converted into a dust
extinction law.  In most cases the resulting R$_V$ is much lower than what
is found in the Milky Way or other galaxies \citep{Tripp:98:815}.  However,
it is clearly not appropriate to assume Milky Way-like dust is responsible
for SN~Ia residual color \citep{Conley:07:L13}.

\begin{figure}[h]
\includegraphics[scale=0.35,angle=90.]{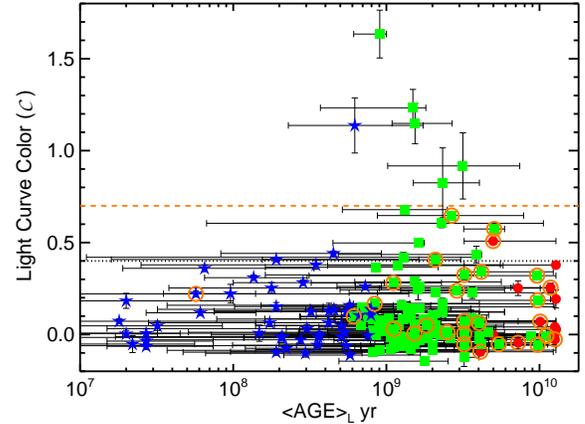}
\caption{Luminosity-weighted age as derived from PEGASE.2 SED 
fits plotted as a function of SN color for the local 
SN~Ia hosts.  The host sSFR is coded as in Figure~\ref{fig_str_ssfr}. The 
SNe with stretch less than $s=0.80$ are overplotted
with open (orange) circles.  The two horizontal lines illustrate two
color cuts applied to our sample: the orange dashed line at $\mathcal{C}=0.7$
is our cut for defining red SNe~Ia, and the black dotted line at
$\mathcal{C}=0.4$ is our cut for deriving Ni masses (see text).  There 
is no obvious correlation, however all of the reddest SNe are hosted by
galaxies with luminosity-weighted ages near 1 Gyr.  It is also clear that
the lowest stretch SNe are not the reddest SNe in our sample.
\label{fig_clr_age}
}
\end{figure}

There is potentially important information about the SN~Ia explosion
contained in the intrinsic color, if we could quantify and remove the
line-of-sight, or host extinction part.  Unfortunately, there are no
observable signatures that allow these two sources to be disentagled,
although extending the light curve data to the near infra-red may minimize
the problem \citep{Kasen:06:939, Vasey:08:377}.  Our host data show no
correlations between host propreties and SN peak color.  The only feature
worth noting, shown in Figure~\ref{fig_clr_age}, is that all of the reddest
SNe appear in hosts with intermediate age ($\sim1$ Gyr).  The low-stretch
SNe with $s < 0.80$, indicated in the plot with the large open (orange)
circles, are not the reddest SNe as might be expected from the
color-luminosity relation \citep{Tripp:98:815, Tripp:99:209,
Parodi:00:634}.  

\begin{figure}[h]
\includegraphics[scale=0.35,angle=90.]{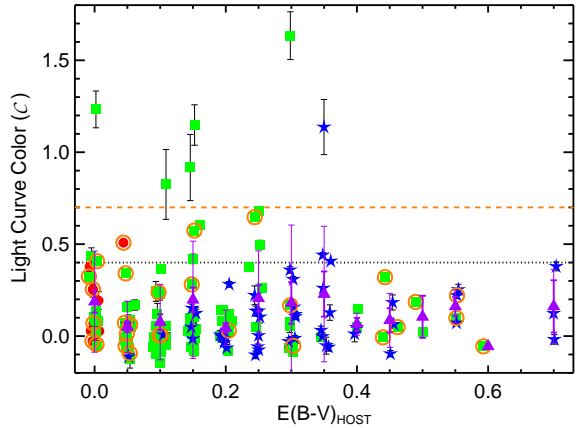}
\caption{SN color as a function of average host color 
excess, $E(B-V)_{HOST}$, as derived from PEGASE.2 SED fits for the
local SN~Ia hosts.  The host sSFR is coded as in Figure~\ref{fig_str_ssfr}.
The SNe with stretch less than $s=0.80$ are overplotted
with open (orange) circles, and the color cut lines are the same as in
Figure~\ref{fig_clr_age}.  To better illustrate the distributions in
each discrete bin of $E(B-V)_{HOST}$ (0.05 Mag), a small random offset was
added to each point.  The average SN color in each host extinction bin is
shown by the filled (purple) triangles.  Note that the sSFR goes up with
host extinction, as expected, and that the reddest SNe appear in hosts with
a range of host extinctions.
\label{fig_clr_ebmv}
}
\end{figure}

Another way to examine this issue is to look at SN color as a function of
host extinction, as shown in Figure~\ref{fig_clr_ebmv}.  We see an average
trend in sSFR with host extinction such that hosts with higher sSFR also
have higher host extinction, a finding that gives us encouragement that the
extinction estimates from PEGASE.2 are robust.  The average SN peak color
(filled purple triangles) shows no trend with host extinction, and the
reddest SNe appear in hosts covering a substantial range of host
extinctions.  This could be due to the clumpyness of host extinction which
makes an average host extinction irrelevant for the specific line-of-sight
to the SN.

\subsection{A Subsample with Low Host Extinction\label{sec_low_extin}}

In an effort to mitigate the difficulty of untangling host extinction and
intrinsic SN color, \citet{Gallagher:08:752} made a study of local SNe~Ia
in hosts that were morphologically classified as early-type and thus
thought to be low-extinction hosts.  They used spectroscopic line
diagnostics to determine metallicities and ages.  We have the opportunity
to select hosts with low extinction inferred from our SED fits, without the
assumption that early-type galaxies all have low extinction and without
possible host classification errors.

To test the impact of host extinction on cosmological fitting, we look for
residual trends in the cosmologically-corrected peak SN~Ia brightness as a
function of host luminosity-weighted age.  We calculate the peak absolute 
B-band brightness for our SNe~Ia, corrected for stretch and color using: 
\begin{equation} M_B
	= m_{Bmax} + \alpha(s-1.0) - \beta\mathcal{C} - \mu_B(z_{spec}),
\end{equation} where $\alpha = 1.2$, $\beta = 2.9$ \citep{Sullivan:09b} and
$\mu_B(z_{spec})$ is the distance modulus based on the spectroscopic
redshift \citep{Conley:09}.  Since we are testing cosmological fitting, we
eliminate low-stretch ($s<0.75$) and red ($\mathcal{C}>0.7$) SNe~Ia.  The
distance modulus, $\mu_B$, is only accurate if the SNe~Ia are in the Hubble
flow so we also eliminate hosts with $z < 0.013$. After applying these
cuts, we are left with a sample of 85 SNe~Ia.

\begin{figure}[h]
\includegraphics[scale=0.35,angle=90.]{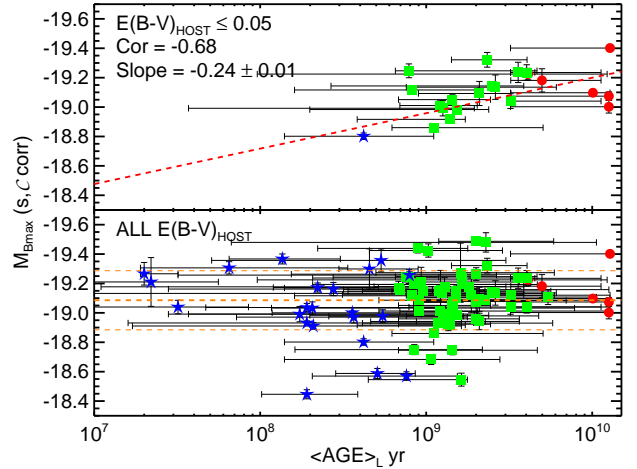}
\caption{SN~Ia stretch-color-corrected M$_B$ as a function of host
luminosity-weighted age for hosts in the Hubble flow ($z > 0.013$) plotted
with the standard symbol coding.  The host sSFR is coded as
in Figure~\ref{fig_str_ssfr}.  Since we are applying stretch and color
corrections, we cut the sample to remove SNe with $s < 0.75$ and
$\mathcal{C}>0.7$.  The {\it bottom} plot shows the Hubble flow subset
(N=85) with no cut on host extinction, while the {\it top} plot shows a low
host extinction subset (N=22) with $E(B-V)_{HOST} \leq 0.05$.  We see no
residual trend with host age for the larger set.  The host extinction cut 
removes the youngest galaxies as expected.  The residual trend in the low 
host extinction subset is indicated by the dashed (red) line in the top
plot.   This trend has a significance of 2.1$\sigma$ when compared with a 
distribution of correlations created when drawing 22 data points at random 
from the total sample 10,000 times.  \label{fig_age_mb}
}
\end{figure}

We plot the stretch-color-corrected absolute peak B magnitude against the
luminosity-weighted host age for this sample in the bottom panel of
Figure~\ref{fig_age_mb} using the previously defined symbols based on sSFR.
In the top panel of Figure~\ref{fig_age_mb}, we plot the same values for a
low host extinction subsample ($N=22$), with $E(B-V)_{HOST} \leq 0.05$
according to our SED fits.  The bottom panel plot shows no obvious residual
trend with age for the larger sample.  The residual trend apparent in the
low host extinction subset has a correlation of $-0.68$.  To determine the
significance of this correlation, we selected 22 hosts at random from the
full set of 85 hosts 10,000 times and ran the same linear regression
analysis on each randomly selected sample.  The low host extinction sample
is a 2.1$\sigma$ outlier when compared with the resulting distribution of
random correlations.  

\subsection{Host Metallicity versus SN~Ia $^{56}$Ni Mass\label{sec_metal_ni}}

Here we complement the higher-redshift ($0.2<z<0.75$) sample of
H09 by using our nearby ($0.013<z<0.06$) sample of SNe~Ia.  This local
sample, in addition to testing the \citet{Timmes:03:L83} model in the
nearby universe, will provide a test of the method developed in H09 for
deriving Ni masses and host metallicities.

\subsubsection{Host Metallicities from Host Masses}

H09 point out the difficulties in using spectroscopic line indices
\citep{Hamuy:00:1479, Gallagher:08:752} or line ratios
\citep{Gallagher:05:210} to derive host metallicities to compare with SN~Ia
intrinsic luminosity or $^{56}$Ni mass.  Instead, they use the
mass-metallicity relation of \citet{Tremonti:04:898}.  There are drawbacks
to using this trend, which uses gas-phase metallicity, on all galaxies
regardless of gas content.  However, as pointed out in H09, there are good
reasons for assuming this method is accurate enough for the purposes here
(see H09, \S3.2).  We also repeat the caution from H09, that gas-phase
metallicity is not the same as stellar metalliticy, although they should be
correlated \citep{Fernandes:05:363}.  The apparent scatter in the Tremonti
relationship is 0.1 dex which we add in quadrature to our errors when we
estimate host metallicities.  For consistency, we use host masses derived
using the same technique applied in H09 and outlined in
\citet{Sullivan:06:868}.

\begin{figure}[h]
\includegraphics[scale=0.35,angle=90.]{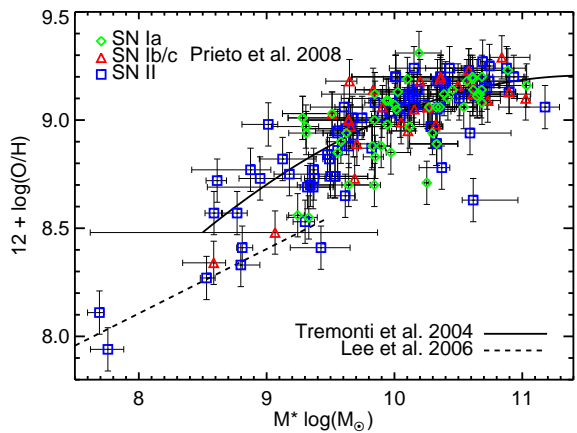}
\caption{Metallicity plotted as a function of PEGASE.2 SED fitted stellar
mass for the hosts from \citet{Prieto:08:999} for which there was sufficient
integrated host photometry to derive good SED fits.  The relation from
\citet{Tremonti:04:898} is indicated as the solid thick line and the
low-metallicity extension from \citet{Lee:06:970} is indicated by the
dashed line.  Open (green)
diamonds indicate hosts of SNe~Ia, open (red) triangles indicate hosts of
SNe~Ib/c, and open (blue) squares indicate hosts of SNe~II.  The Prieto
sample is poorly sampled at masses below $\sim
10^{9.2}$ M$_{\odot}$.  This is most likely due to selection effects from
host-targeted local SN surveys that rarely target low-mass hosts.
Most of our local well fit SNe are in hosts with M$_* \gtrsim 10^9$ 
M$_{\odot}$ (see Figure~\ref{fig_str_mass}), thus we should not be overly 
influenced by this selection effect.
\label{fig_mass_metal}
}
\end{figure}

We test the use of the \citet{Tremonti:04:898} mass-metallicity relation
for low-z SN hosts by using the SN host metallicity data from
\citet{Prieto:08:999}.  Their metallicities come from the SDSS as were
those used in \citet{Tremonti:04:898}.  A comparison of the two results
should reveal any biases in the low-z SN host sample.  We have sufficiently
good photometry for 170 SN hosts in the \citet{Prieto:08:999} catalog to
perform this comparison.  These data are also from the GLGA and will be
published in \citet{Seibert:09}.  We do not place any requirements on the
SNe in this sample other than that they have metallicities from
\citet{Prieto:08:999} and that we have enough host photometry for a good
SED fit.  Figure~\ref{fig_mass_metal} plots our SED-fit stellar masses
against the metallicities from \citet{Prieto:08:999}.  The relation from
\citet{Tremonti:04:898} is shown as the solid line and the low-metallicity
extension from \citet{Lee:06:970} is shown as the dashed line for
comparison with the host data.  The hosts with masses in the range $9 <
\log$ M$_* < 11$ appear to follow the Tremonti relation well, while hosts
with $\log$ M$_* < 9$ appear to follow the Lee relation, albeit with a
sparser sampling.

We will be looking for average trends in $^{56}$Ni mass with metallicity,
therefore we want to be sure to sample the range of the scatter in the
Tremonti relation so we are not biased toward one side or the other by
observational effects.  The host mass range for our low-z SNe~Ia with
well-fit light curves is primarily above $\log$ M$_* = 9$ (see
Figure~\ref{fig_str_mass}).  The scatter in the mass-metallicity relation
for SN hosts appears to be well sampled above this mass, thus we should
be less subject to observational biases that could be in effect below
this host mass.  We also need not use the low-metallicity extension from
\citet{Lee:06:970} for our base sample of SN~Ia hosts.

\subsubsection{SN~Ia $^{56}$Ni Masses}

H09 present a technique for estimating the SN~Ia $^{56}$Ni mass using
Arnett's Rule \citep{Arnett:79:L37, Arnett:82:785} to convert bolometric
luminosity and SN~Ia rise time to $^{56}$Ni mass.  We use the identical
technique on the well-sampled light curves for our local SNe~Ia.  Because
the estimate of the bolometric luminosity assumes the luminosity distance
is accurate, we use only the SNe in the Hubble flow (see Table~\ref{tab_sne}).  

The lowest stretch SNe~Ia are also excluded from this experiment because
they produce inaccurate $^{56}$Ni masses using our technique.  As
previously stated, low luminosity SNe~Ia deviate from the normal trend of
luminosity versus light-curve shape
\citep{Garnavich:04:1120,Taubenberger:08:75}, thus the bolometric
luminosity is inaccurate.  The SN~Ia spectral templates currently available
are also not representative of SNe~Ia with $s < 0.75$
\citep{Conley:08:482}.  A higher-order relation between light-curve shape
and luminosity that properly accounts for low-luminosity SNe~Ia and better
spectral templates would allow us to use these SNe to test this model
\citep{Gaitan:09}.  They will be quite useful because the strongest effect
from metallicity should occur in the faintest SNe.  For our plots of Ni
mass, we exclude objects with $s < 0.75$.  To preclude inaccuracies due to
excessive extinction we include only SNe~Ia with $\mathcal{C} < 0.4$.  The
final sample of 74 SNe that pass these cuts is presented in
Table~\ref{tab_ni_metal}, where the host, and the derived $^{56}$Ni masses
and host metallicities are listed.

\begin{figure}[h]
\includegraphics[scale=0.35,angle=90.]{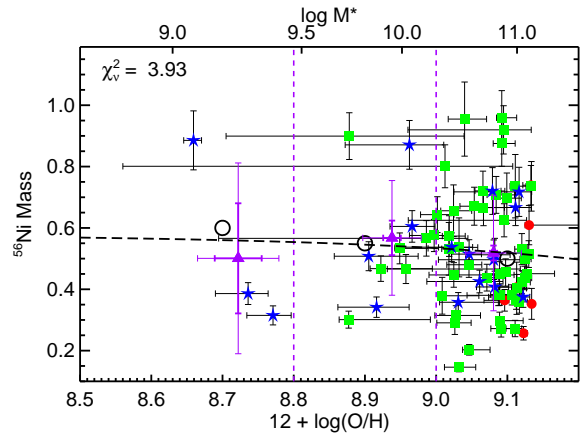}
\caption{SN~Ia  $^{56}$Ni mass as a function of metallicity.  The host sSFR
is coded as in Figure~\ref{fig_str_ssfr}.  The average $^{56}$Ni masses in
(O/H) bins of 0.2 dex starting at $12 + \log$(O/H)$ = 8.6$ (vertical dashed
purple lines) are plotted as
filled (purple) triangles.  The thin errorbars are the RMS in each bin,
while the thick errorbars are the errors in the mean within the bin.  The
data from H09 is shown as the large open (black) circles.  The expected
trend with metallicity from the \citet{Timmes:03:L83} model, altered for
thin disk O/Fe, is plotted as the thick dashed line.  The average trend is
consistent with that predicted by \citet{Timmes:03:L83} as was found by H09
for an intermediate redshift sample of SNe~Ia ($0.2 < z < 0.75$).
\label{fig_ni_metal}
}
\end{figure}

Figure~\ref{fig_ni_metal} shows the SN~Ia $^{56}$Ni mass as a function of
host metallicity for our final sample.  We use the same symbol coding as in
Figure~\ref{fig_str_ssfr}.  We have taken the average Ni mass in three bins
of metallicity starting at $12 + \log$ (O/H) $ = 8.6$ and having a width of
0.2 and plot these as filled (purple) triangles.  The model from
\citet{Timmes:03:L83} altered for thin disk O/Fe (H09) is overplotted as a
thick dashed line.  The average trend seen in H09 is shown by the open
circles.  Although our data are statistically consistent with no trend,
they are also consistent with the H09 trend and the Timmes model.

\subsubsection{Color Correction}

\begin{figure}[h]
\includegraphics[scale=0.35,angle=90.]{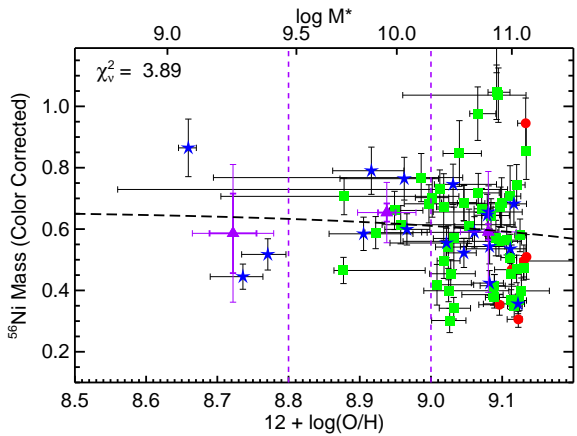}
\caption{SN~Ia  $^{56}$Ni mass as a function of metallicity with a
correction for SN color applied (see text).  The host sSFR is coded as
in Figure~\ref{fig_str_ssfr}.  The Timmes model is plotted
as the thick dashed line.  The color correction was not as effective as
in the H09 sample and only reduces the $\chi^2_{\nu}$ from 3.93 to 3.89, when
comparing the data to the Timmes model.
\label{fig_ni_metal_cc}
}
\end{figure}

H09 apply a correction for the intrinsic SN~Ia light curve color to their
derived Ni masses, after a color cut has been applied.  Fainter SNe~Ia are
redder due to the intrinsic color-luminosity relation, or line-of-sight
extinction, or both.  Our Ni mass calculation requires the intrinsic
luminosity, corrected for the dimming effects of dust, but not corrected
for color-luminosity.  Since the correction is based on the observed and
not intrinsic color of the SN there is bound to be some error in assuming
that there is no dust dimming (uncorrected Ni-masses) or that the
color-luminosity relation is purely due to dust (color corrected
Ni-masses).  Both will be inaccurate at some level.  One way to examine
this is to calculate and compare the $\chi^2/DOF$ or $\chi^2_{\nu}$ that
results when comparing the corrected and uncorrected data to the Timmes
model.  When we apply the color correction, we find a mild improvement in
the scatter as shown in Figure~\ref{fig_ni_metal_cc}.  When compared with
the Timmes model, the $\chi^2_{\nu}$ is only marginally improved going from
3.93 (uncorrected) to 3.89 (color corrected).

These values of $\chi^2_{\nu}$ greater than one imply that our errors do
not reflect the expected scatter of the data points around the Timmes
model, assuming it is correct.  H09 found that in order to achieve
$\chi^2_{\nu} = 1$, they had to add in quadrature an extra 0.16~M$_{\odot}$
to their errors in Ni mass for their uncorrected masses.  We also find we
need an extra error of 0.16 M$_{\odot}$ in our uncorrected $^{56}$Ni masses
to achieve $\chi^2_{\nu} = 1$.  Another measure of the effectiveness of the
color correction is to see how much this extra error is reduced by the
correction.  For H09 it was significantly reduced from 0.16 (uncorrected)
to 0.12 (color corrected) M$_{\odot}$.  For our data there was only a
marginal decrease in the extra scatter required for the color corrected
$^{56}$Ni masses (0.162 to 0.158 M$_{\odot}$).

One possible explanation of this difference is the homogeneous light curve
photometry available for the H09 sample in contrast to the heterogeneous
photometry available for our sample.  The color cuts in H09 were made in a
multi-color space that included wavelength regions that were not available
for all the SNe in the local sample (see their \S3.3).  Consequently, our
color cut here was in a single color (B-V) and may not have been as
effective at removing outliers in the color corrected $^{56}$Ni mass.  We
point out that there is an obvious reduction by the color correction in the
scatter of Ni mass for lower metallicity hosts and the biggest outliers in
the color-corrected plot (Figure~\ref{fig_ni_metal_cc}) are near the
highest metallicity.  The fact that the color correction provides any
improvement at all is consistent with the idea that at least some part of
the color-luminosity relation is due to host dust extinction.

\subsubsection{SN~Ia $^{56}$Ni Mass in Three Stretch Bins}

\begin{figure}[h]
\includegraphics[scale=0.35,angle=90.]{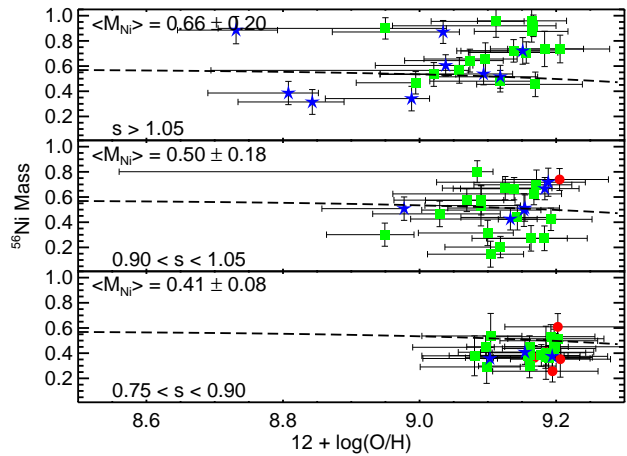}
\caption{SN~Ia $^{56}$Ni mass (uncorrected) as a function of metallicity
for three stretch bins, as annotated.  The host sSFR is coded as
in Figure~\ref{fig_str_ssfr}.  The Timmes model is plotted as
the thick dashed line.  The average $^{56}$Ni mass increases with stretch
across the three bins, as expected.  The range of host metallicity and
the scatter in Ni mass decreases toward lower stretch as was seen in the 
higher redshift sample from H09.
\label{fig_ni_metal_slice}
}
\end{figure}

Figure~\ref{fig_ni_metal_slice} shows the uncorrected Ni mass versus
metallicity divided in three stretch bins (see H09, Figure~10).  We do not
sample as large a range of metallicity as H09, nonetheless, we do see
similar features.  The range of host metallicity for the higher stretch
SNe~Ia is roughly double that for the lower stretch group.  We expect the
average $^{56}$Ni mass to increase with stretch due to stretch-luminosity,
but we also see that the scatter in the masses is lowest in the lowest
stretch group.  The stretch cut in the low-stretch bin undoubtedly limits
the Ni mass in this bin (see H09, \S4.5), but at the low Ni-mass end our
local sample should be less subject to Malquist biases.  Another effect is
the color cut of $\mathcal{C} < 0.4$, which precludes the reddest,
faintest and therefore lowest Ni-mass SNe.  Nonetheless, the trends are
suggestive and support the idea that lower-stretch SNe~Ia appear in higher
metallicity hosts.

\subsection{Host Age versus SN~Ia $^{56}$Ni Mass\label{sec_age_ni}}

\begin{figure}[h]
\includegraphics[scale=0.35,angle=90.]{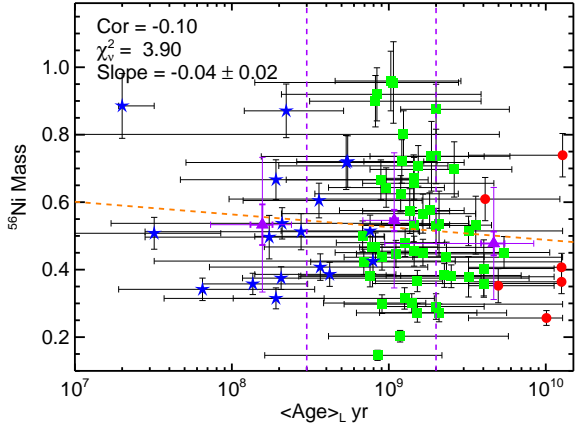}
\caption{SN~Ia  $^{56}$Ni mass, with no color correction, as a function 
of luminosity weighted host age.  The host sSFR is coded as
in Figure~\ref{fig_str_ssfr}.  The average $^{56}$Ni mass in three bins
dividing the sample at luminosity-weighted ages of $3 \times 10^8$ and
$2 \times 10^9$ yr (vertical dashed purple lines) are plotted as filled
purple triangles.  The thin errorbars are the RMS in each bin while the
thick errorbars are the error in the mean in each bin.  A linear fit is
shown as the dashed orange line and the results of the correlation are
annoted on the plot.  The correlation is weaker and the slope
is shallower than what was seen in the H09 sample (see their Figure~8).  
The linear trend with age produces a $\chi^2_{\nu}$ similar to the Timmes 
model.  Combined with H09, Figure~8, we note that SNe~Ia with
$^{56}$Ni masses greater than $\sim0.5$ M$_{\odot}$ are rare in hosts 
with luminosity-weighted ages greater than $\sim3\times10^9$ yr.
\label{fig_ni_age}
}
\end{figure}

\begin{figure}[ht]
\includegraphics[scale=0.35,angle=90.]{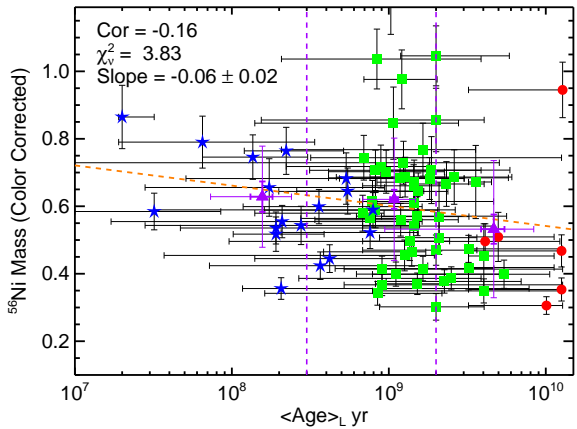}
\caption{SN~Ia  $^{56}$Ni mass, corrected for SN color, as a function 
of luminosity weighted host age.  The host sSFR is coded as
in Figure~\ref{fig_str_ssfr} and the bins for the averages are the same as
in Figure~\ref{fig_ni_age}.  The color correction increases the
correlation and marginally changes the slope.  The rarity of higher Ni-mass
SNe~Ia in the oldest hosts is still apparent (see Figure~\ref{fig_ni_age} 
and H09, Figure~9).
\label{fig_ni_age_cc}
}
\end{figure}

Due to the age-metallicity degeneracy, we present plots of $^{56}$Ni mass
against luminosity-weighted age for the uncorrected
(Figure~\ref{fig_ni_age}) and color corrected (Figure~\ref{fig_ni_age_cc})
Ni masses.  We use the same symbol scheme as previous plots.  We use a
linear regression analysis to measure and test the correlation between age
and Ni mass.  The color correction slightly steepens the slope of the fit
(from $-0.04\pm0.02$ to $-0.06\pm0.02$), and does improve the correlation
from $-0.10$ (uncorrected) to $-0.16$ (color corrected).

The age-Ni mass slopes reported in H09 for their sample are significantly
steeper: $-0.15\pm0.03$ (uncorrected) and $-0.11\pm0.03$ (color corrected).
The correlations are also larger: $-0.38$ (uncorrected) and $-0.37$ (color
corrected).  Examining Figures~8 and 9 in H09, we see that a large part of
the correlation seems to originate in hosts with luminosity-weighted ages
beyond 3 Gyr which seem incapable of producing SNe with $^{56}$Ni masses
above 0.5 M$_{\odot}$ in the uncorrected plot (their Figure~8) and above
0.6 M$_{\odot}$ in the corrected plot (their Figure~9).  Below 3 Gyr,
there appears to be very little correlation between age and Ni
mass.  We see a similar feature in our data as well
(Figures~\ref{fig_ni_age} and \ref{fig_ni_age_cc}), although not quite as
strong.  This feature implies a threshhold that is reached at a
luminosity-weighted age of $\sim$ 3 Gyr, rather than a uniform trend with
age.

\section{Discussion\label{sec_discuss}}

The implications of these results must be tempered by the limitations of
the techniques.  In particular, the SED-fitting method employed was
optimized for higher redshift galaxies and we do not employ infra-red data
to constrain the SEDs of our low-redshift hosts.  Our aim is to compare
hosts at high and low redshifts using the same methods.  We focus here on
trends and this comparison and will pursue more precise determinations of host
parameters with infra-red data and more detailed methods in a future paper.

We reiterate that the ages estimated with the PEGASE.2 fits are luminosity 
weighted and the masses are current stellar masses.  The star
formation rates only measure current activity based on UV and optical data
and have not been modified by FIR data to account for dust.  The global
host extinctions are derived by extincting the entire SED and do not
account for variable thermal components in the IR.

\subsection{Host Properties at High and Low Redshift}

Our results show features similar to those presented in higher redshift
host studies using the same methods \citep[][H09]{Sullivan:06:868,
Sullivan:09:539}.  Low stretch SNe~Ia appear to be rare in high sSFR hosts
and the average stretch of SNe~Ia in the lowest sSFR hosts is smaller than
in hosts with higher sSFR (see Figure~\ref{fig_str_ssfr}).  The average
stretch of SNe in hosts younger than $\sim$2 Gyr is close to 1 and for
older hosts the average stretch appears to drop below 1 (see
Figure~\ref{fig_str_age}).  H09 were the first to plot host stellar mass
against stretch, but their sample does not include SNe~Ia with $s < 0.75$
(see their Figure 4).  The interesting feature in our plot
(Figure~\ref{fig_str_mass}) is the limited mass range for lower-stretch
SNe.  This feature is present in the H09 figure, but it is less pronounced
due to having fewer low-stretch SNe in their sample.

The apparent stellar mass threshhold of $10^{10}$ M$_{\odot}$ for the hosts
of low-luminosity SNe~Ia is fairly robust against selection effects.  Our
sample includes many lower-mass hosts within which it is easier to detect
low-luminosity SNe.  That these SNe prefer higher-mass hosts supports the
idea that they arise from an older progenitor population.
Figures~\ref{fig_str_ssfr}, \ref{fig_str_age} and \ref{fig_str_mass} show
that the one high sSFR host (E445-G66) of a SNe~Ia with $s < 0.80$
(SN1993H) is also massive, implying that the recently formed populations
are mixed with the older populations that could be responsible for the
low-stretch SN~Ia.  The correlation between mass and metalicity implies
that metallicity could play a role as well (see \S\ref{sec_metal_ni}).  We
see no low-mass, high sSFR hosts of low-stretch SNe~Ia and we also see few
older, higher-mass hosts of the highest stretch SNe~Ia.  A method that can
more accurately asses the star formation history of these hosts and the
relative contributions from older and younger populations will produce a
more definitive constraint on the progenitors of the low-luminosity SNe~Ia
\citep{Schawinski:09}.

Our plot of SN peak color versus host age illustrates the difficulty of
disentangling the source of the color (see Figure~\ref{fig_clr_age}) since
it shows no correlation.  We are also challenged in our comparison with
higher redshift samples by the bias against SNe with very red peak colors
intrinsic to cosmology surveys (see \S\ref{sec_sample}).  Even in the local
universe, our sample of six SNe~Ia with $\mathcal{C}>0.7$ is too small for
definitive interpretation.  We point out one common feature of our figure and
Figure~3 in \citet{Sullivan:09b}: very red SNe appear to be rare in
hosts less than 0.5 Gyr in luminosity-weighted age.

Based on Figure~\ref{fig_ni_metal}, we conclude that the model of
\citet{Timmes:03:L83}, relating metallicity to produced $^{56}$Ni mass,
appears consistent with the derived host metallicities and SN~Ia $^{56}$Ni
masses for SNe in the local universe as well as at higher redshifts (H09),
although the local data are also consistent with no trend in these
properties.  The degeneracy between age and metallicity makes it difficult
to decide which is the more important factor in determining the $^{56}$Ni
mass of hosted SNe~Ia.  The apparent transition at a luminosity-weighted
host age of 3 Gyr (see Figure~\ref{fig_ni_age} and H09, Figure 8) could
place interesting constraints on progenitor scenarios given that this
implies a fairly long delay time for low $^{56}$Ni mass, hence
low-luminosity and low-stretch SNe~Ia.

\subsection{Host Properties and Cosmological Fitting}

It is encouraging that the features in plots of light curve properties
against host properties in local hosts are similar when compared with the
SNLS sample out to $z=0.75$ \citep{Astier:06:31, Howell:07:L37,
Sullivan:09b}, although a detailed comparison using Hubble diagram
residuals will be needed to determine if any significant evolution of the
population can be detected \citep{Conley:09, Sullivan:09b}.  These trends
do signal potential systematics for SN cosmology: as we extend the samples
beyond redshift $z=0.75$, we know the mix of host populations will move
toward lower-mass, higher sSFR galaxies.

Our examination of low-extinction hosts suggests a trend in brightness with
host age such that SNe~Ia in older hosts appear brighter after correcting
for stretch and color (see Figure~\ref{fig_age_mb}).  While the correlation
we see is hardly significant (2.1$\sigma$), it does suggest that
luminosity-weighted host age might be an interesting parameter to explore
in cosmological fits of SNe~Ia.  A larger sample of SNe~Ia from
low-extinction hosts would allow a more definitive test of this
correlation.

While there is no apparent trend in a similar plot in \citet[see their
Figure~9]{Gallagher:08:752}, we see a trend that is marginally significant.
There are several possible explanations for this difference.  Since the
hosts in the \citet{Gallagher:08:752} sample were selected morphologically,
their sample may not be as free from dust extinction as originally thought.
In addition, their corrections for light curve shape are derived with a
method that assumes a Milky Way-like dust is responsible for residual color
after a quadratic color-decline relation is subtracted \citep[see
comparison of fitting methods in][]{Conley:08:482}, which may not be
appropriate for these hosts.  The trend we see in Figure~\ref{fig_age_mb}
is in the right sense to be caused by residual line-of-sight extinction
since we would expect the SNe hosted by the oldest galaxies to have the
least non-local extinction.

These data suggest that either the SNe~Ia or the dust produced in galaxies
of different host properties are different or both.  \citet{Hicken:09a}
showed a weak trend in the Hubble residuals when using samples divided into
three host morphology bins, after removing the reddest SNe.  An exploration
based on more physical host properties could provide a more accurate
assessment of this effect \citep{Sullivan:09b}.  Further progress may be
made by extending host photometry farther into the infra-red to help
constrain the host dust content and by more detailed fitting of host
star-formation history to constrain the most recent episode of star
formation \citep[e.g.,][]{Schawinski:09}.

\section{Summary\label{sec_sum}}

We gathered the host integrated photometry for 168 well-observed
SNe~Ia and correlated the derived host properties with the hosted SN~Ia
light curve properties.  We used the derived average host extinction to
isolate a set of SNe~Ia arising in hosts with low extinction.  We used
the methods outlined in H09 to compare host metallicity with SN~Ia
$^{56}$Ni mass.

With our low-redshift sample, we have corroborated trends in SN~Ia light
curve shape with host properties observed in higher-redshift SN~Ia samples
of similar size \citep{Sullivan:06:868}.  Specifically, we find that the
higher the sSFR of the host galaxy, the brighter and slower the SNe~Ia that
are produced, on average.  We find that the typical stretch of the SNe~Ia
produced begins to drop after the host galaxy reaches a luminosity-weighted
age of 1 Gyr.  We find that low stretch ($s < 0.80$) SNe~Ia in the local
universe prefer hosts with stellar masses above $10^{10}$
M$_{\odot}$.  This is unlikely to be due to a Malmquist bias, since lower
luminosity SNe are easier to detect against lower mass, lower luminosity
hosts.  We find that fainter, faster SNe~Ia do indeed prefer more massive
and presumably higher metallicity hosts, supporting the link between the
production of $^{56}$Ni and progenitor metallicity.

We suggest that it is likely that SN~Ia cosmological fitting could be
improved by adding a parameter characterizing the host properties to
account for the impact of galaxy evolution on the evolution of SNe~Ia with
redshift.  The trends in host properties with SN~Ia light curve properties
support this idea and suggest that perhaps host age would be a good
property to use.

Our sample shows no obvious trend in host properties with SN~Ia peak color,
except that all the reddest SNe~Ia appear in hosts with ages around 1 Gyr.
We find that the reddest SNe~Ia appear in hosts with a range of
derived host extinctions.

Our low host extinction sample suggests a residual trend of SN peak
absolute brightness and host age such that SNe~Ia in older hosts appear
brighter after their peak brightnesses are corrected for light-curve shape
and color (i.e.  stretch-color corrected).  One possible explanation for
this is that the color correction leaves a small residual brightness trend
that is due to extinction and that older hosts contain less dust than
younger hosts.  This residual trend is consistent with the one reported in
\citet{Hicken:09a}, who find that SNe~Ia in early-type galaxies are
brighter, after shape and color correction, than SNe~Ia appearing in
late-type hosts.

For SNe~Ia with $s > 0.75$ and $\mathcal{C} < 0.4$, we find local SNe~Ia
host metallicities and SN~Ia $^{56}$Ni masses are consistent with the model
of \citet{Timmes:03:L83} altered for thin disk O/Fe as in H09, but with an
additional scatter of 0.16 M$_{\odot}$.  The local data are also consistent
with no trend.  The failure of the SN color correction to significantly
reduce the scatter in the Ni-masses is likely due to the heterogeneous
photometry of the local SNe~Ia.  In spite of this, the trend of average
$^{56}$Ni mass and the scatter with stretch is very similar to that seen in
H09.

We examined the trend in $^{56}$Ni mass with host luminosity-weighted age
and found a shallower slope and a weaker correlation than shown in H09 for
a higher redshift sample ($0.2 < z < 0.75$).  We point out what appears to
be a threshhold luminosity-weighted age of 3 Gyr apparent in our sample and
the H09 sample, above which a host is less likely to produce SNe~Ia with
$^{56}$Ni masses greater than $\sim 0.5$ M$_{\odot}$.  Below this age
threshhold, there appears to be little correlation between $^{56}$Ni mass
and host age.



\acknowledgments

J. N. would like to thank Fillipo Mannucci, Dan Maoz, Massimo Della Valle,
and Patrizia Braschi, the organizers of the 2008 May SN~Ia rates conference
in Florence, Italy where a preliminary version of this work was presented
and discussed.  We acknowledge the useful comments by the anonymous
referee.

GALEX (Galaxy Evolution Explorer) is a NASA Small Explorer, launched in
2003 April. We gratefully acknowledge NASA's support for construction,
operation, and science analysis for the GALEX mission, developed in
cooperation with the Centre National d'Etudes Spatiales of France and the
Korean Ministry of Science and Technology.

This research has made use of the NASA/IPAC Extragalactic Database (NED)
which is operated by the Jet Propulsion Laboratory, California Institute of
Technology, under contract with the National Aeronautics and Space
Administration.

Funding for the SDSS and SDSS-II has been provided by the Alfred P.
Sloan Foundation, the Participating Institutions, the National Science
Foundation, the U.S. Department of Energy, the National Aeronautics and
Space Administration, the Japanese Monbukagakusho, the Max Planck
Society, and the Higher Education Funding Council for England. The SDSS
Web Site is http://www.sdss.org/.

The SDSS is managed by the Astrophysical Research Consortium for
the Participating Institutions. The Participating Institutions are
the American Museum of Natural History, Astrophysical Institute
Potsdam, University of Basel, University of Cambridge, Case Western
Reserve University, University of Chicago, Drexel University,
Fermilab, the Institute for Advanced Study, the Japan Participation
Group, Johns Hopkins University, the Joint Institute for Nuclear
Astrophysics, the Kavli Institute for Particle Astrophysics and
Cosmology, the Korean Scientist Group, the Chinese Academy of
Sciences (LAMOST), Los Alamos National Laboratory, the
Max-Planck-Institute for Astronomy (MPIA), the Max-Planck-Institute
for Astrophysics (MPA), New Mexico State University, Ohio State
University, University of Pittsburgh, University of Portsmouth,
Princeton University, the United States Naval Observatory, and the
University of Washington.

\renewcommand{\arraystretch}{1.15}

\clearpage

\begin{longtable}{llcrrrrr}
\tablecaption{SN Ia Sample\label{tab_sne}}
\tablewidth{0pt}
\tablehead{
& &
\colhead{$cz$} \\
\colhead{SN} & \colhead{HOST} &
\colhead{($\log$ km s$^{-1}$)} &
\colhead{stretch (s)\tablenotemark{1}} &
\colhead{$m_{Bmax}$} &
\colhead{Status\tablenotemark{2}}
}
1980N    & NGC1316          &   3.26 &   0.83$\pm$0.01 &  12.42$\pm$0.01 &  \\
1981B    & NGC4536          &   3.26 &   0.88$\pm$0.02 &  11.95$\pm$0.01 & C \\
1981D    & NGC1316          &   3.26 &   0.86$\pm$0.04 &  12.54$\pm$0.05 & C \\
1990N    & NGC4639          &   3.00 &   1.09$\pm$0.01 &  12.70$\pm$0.02 & C \\
1991T    & NGC4527          &   3.24 &   1.07$\pm$0.02 &  11.44$\pm$0.03 &  \\
1991U    & IC4232           &   3.98 &   1.01$\pm$0.05 &  16.33$\pm$0.08 & H, C \\
1991ag   & IC4919           &   3.64 &   1.10$\pm$0.03 &  14.43$\pm$0.08 & H, C \\
1992A    & NGC1380          &   3.27 &   0.81$\pm$0.01 &  12.54$\pm$0.01 & C \\
1992P    & IC3690           &   3.88 &   1.16$\pm$0.12 &  16.05$\pm$0.03 & H, C \\
1992ag   & E508-G67         &   3.88 &   0.98$\pm$0.05 &  16.27$\pm$0.05 & H, C \\
1992bc   & E300-G09         &   3.76 &   1.07$\pm$0.01 &  15.10$\pm$0.01 & H, C \\
1992bl   & E291-G11         &   4.12 &   0.78$\pm$0.03 &  17.31$\pm$0.05 & L, H, C \\
1992bo   & E352-G57         &   3.75 &   0.75$\pm$0.01 &  15.73$\pm$0.02 & L, H, C \\
1993H    & E445-G66         &   3.86 &   0.66$\pm$0.06 &  16.79$\pm$0.03 & L, H, C \\
1993ae   & UGC1071          &   3.76 &   0.77$\pm$0.03 &  16.24$\pm$0.05 & L, H \\
1994M    & NGC4493          &   3.84 &   0.80$\pm$0.03 &  16.21$\pm$0.04 & H, C \\
1994Q    & PGC0059076       &   3.95 &   1.07$\pm$0.07 &  16.36$\pm$0.08 & H \\
1994S    & NGC4495          &   3.66 &   1.04$\pm$0.04 &  14.78$\pm$0.02 & H, C \\
1994ae   & NGC3370          &   3.11 &   1.05$\pm$0.01 &  12.95$\pm$0.02 & C \\
1995D    & NGC2962          &   3.29 &   1.11$\pm$0.01 &  13.26$\pm$0.03 & C \\
1995E    & NGC2441          &   3.54 &   0.93$\pm$0.03 &  16.69$\pm$0.02 & C \\
1995al   & NGC3021          &   3.19 &   1.08$\pm$0.03 &  13.32$\pm$0.02 & C \\
1995bd   & UGC3151          &   3.66 &   1.03$\pm$0.01 &  15.28$\pm$0.20 & H, C \\
1996C    & M+08-25-47       &   3.91 &   1.07$\pm$0.04 &  16.63$\pm$0.03 & H, C \\
1996X    & NGC5061          &   3.31 &   0.85$\pm$0.02 &  12.99$\pm$0.03 & C \\
1996Z    & NGC2935          &   3.36 &   0.92$\pm$0.08 &  14.32$\pm$0.08 & C \\
1996ai   & NGC5005          &   3.00 &   1.11$\pm$0.03 &  16.90$\pm$0.01 & R, C \\
1996bo   & NGC673           &   3.72 &   0.87$\pm$0.01 &  15.84$\pm$0.03 & H, C \\
1996bv   & UGC3432          &   3.70 &   1.05$\pm$0.05 &  15.32$\pm$0.05 & H, C \\
1997E    & NGC2258          &   3.60 &   0.81$\pm$0.02 &  15.11$\pm$0.05 & C \\
1997Y    & NGC4675          &   3.68 &   0.91$\pm$0.07 &  15.28$\pm$0.07 & H \\
1997bp   & NGC4680          &   3.40 &   0.97$\pm$0.02 &  13.91$\pm$0.02 & C \\
1997bq   & NGC3147          &   3.45 &   0.88$\pm$0.02 &  14.33$\pm$0.04 &  \\
1997br   & E576-G40         &   3.32 &   0.91$\pm$0.04 &  13.31$\pm$0.08 &  \\
1997cw   & NGC105           &   3.72 &   1.13$\pm$0.04 &  15.99$\pm$0.06 & H, C \\
1997do   & UGC3845          &   3.48 &   0.94$\pm$0.03 &  14.27$\pm$0.04 & C \\
1998V    & NGC6627          &   3.72 &   1.00$\pm$0.04 &  15.08$\pm$0.08 & H, C \\
1998ab   & NGC4704          &   3.91 &   0.94$\pm$0.02 &  16.06$\pm$0.03 & H, C \\
1998aq   & NGC3982          &   2.97 &   0.95$\pm$0.01 &  12.31$\pm$0.01 & C \\
1998bu   & NGC3368          &   2.97 &   0.95$\pm$0.02 &  12.10$\pm$0.01 & C \\
1998de   & NGC252           &   3.70 &   0.57$\pm$0.02 &  17.36$\pm$0.03 & L, H, C \\
1998dh   & NGC7541          &   3.43 &   0.91$\pm$0.02 &  13.86$\pm$0.04 &  \\
1998dm   & M-01-04-44       &   3.29 &   1.07$\pm$0.07 &  14.64$\pm$0.08 & C \\
1998dx   & UGC11149         &   4.21 &   0.80$\pm$0.04 &  17.55$\pm$0.04 & L, H, C \\
1998ec   & UGC3576          &   3.77 &   0.98$\pm$0.07 &  16.17$\pm$0.11 & H \\
1998eg   & M+01-57-14       &   3.87 &   0.92$\pm$0.06 &  16.11$\pm$0.05 & H \\
1998es   & NGC632           &   3.50 &   1.14$\pm$0.01 &  13.84$\pm$0.02 & C \\
1999X    & CGCG180-22       &   3.88 &   0.91$\pm$0.08 &  16.09$\pm$0.13 & H, C \\
1999aa   & NGC2595          &   3.65 &   1.13$\pm$0.01 &  14.73$\pm$0.02 & H, C \\
1999ac   & NGC6063          &   3.45 &   0.98$\pm$0.01 &  14.12$\pm$0.02 & C \\
1999cc   & NGC6038          &   3.97 &   0.78$\pm$0.02 &  16.78$\pm$0.01 & L, H, C \\
1999cl   & NGC4501          &   3.33 &   0.96$\pm$0.02 &  14.87$\pm$0.02 & R, C \\
1999cp   & NGC5468          &   3.45 &   1.01$\pm$0.03 &  13.95$\pm$0.02 & C \\
1999da   & NGC6411          &   3.57 &   0.55$\pm$0.02 &  16.60$\pm$0.03 & L, C \\
1999dk   & UGC1087          &   3.66 &   0.97$\pm$0.03 &  14.83$\pm$0.03 & H, C \\
1999dq   & NGC976           &   3.64 &   1.12$\pm$0.01 &  14.42$\pm$0.05 & H, C \\
1999ee   & IC5179           &   3.53 &   1.07$\pm$0.01 &  14.85$\pm$0.01 & C \\
1999gd   & NGC2623          &   3.74 &   0.92$\pm$0.06 &  16.93$\pm$0.03 & H, C \\
2000E    & NGC6951          &   3.12 &   1.06$\pm$0.01 &  12.85$\pm$0.15 & C \\
2000ca   & E383-G32         &   3.86 &   1.08$\pm$0.03 &  15.57$\pm$0.03 & H, C \\
2000ce   & UGC4195          &   3.69 &   1.02$\pm$0.03 &  17.06$\pm$0.04 & H, C \\
2000cx   & NGC524           &   3.38 &   0.87$\pm$0.01 &  13.06$\pm$0.04 &  \\
2000dk   & NGC382           &   3.72 &   0.74$\pm$0.01 &  15.36$\pm$0.03 & L, H, C \\
2000fa   & UGC3770          &   3.80 &   1.03$\pm$0.03 &  15.86$\pm$0.05 & H, C \\
2001N    & NGC3327          &   3.80 &   0.94$\pm$0.05 &  16.58$\pm$0.04 & H, C \\
2001V    & NGC3987          &   3.66 &   1.13$\pm$0.01 &  14.62$\pm$0.02 & H, C \\
2001ay   & IC4423           &   3.96 &   1.59$\pm$0.03 &  16.77$\pm$0.03 & H \\
2001az   & UGC10483         &   4.09 &   1.08$\pm$0.06 &  16.93$\pm$0.04 & H, C \\
2001ba   & M-05-28-01       &   3.95 &   1.01$\pm$0.02 &  16.21$\pm$0.03 & H, C \\
2001da   & NGC7780          &   3.71 &   0.89$\pm$0.24 &  15.48$\pm$0.09 & H, C \\
2001el   & NGC1448          &   3.07 &   0.98$\pm$0.01 &  12.78$\pm$0.01 & C \\
2001en   & NGC523           &   3.69 &   0.90$\pm$0.05 &  15.05$\pm$0.08 & H, C \\
2001ep   & NGC1699          &   3.59 &   0.88$\pm$0.02 &  14.90$\pm$0.02 & C \\
2001fe   & UGC5129          &   3.61 &   1.09$\pm$0.03 &  14.70$\pm$0.02 & H, C \\
2001ie   & UGC5542          &   3.97 &   0.80$\pm$0.05 &  16.78$\pm$0.05 & L, H, C \\
2002bf   & CGCG266-031      &   3.86 &   0.94$\pm$0.03 &  16.33$\pm$0.05 & H, C \\
2002bo   & NGC3190          &   3.11 &   0.94$\pm$0.01 &  13.96$\pm$0.02 & C \\
2002cd   & NGC6916          &   3.49 &   1.07$\pm$0.02 &  15.50$\pm$0.16 & C \\
2002ck   & UGC10030         &   3.97 &   1.06$\pm$0.06 &  16.28$\pm$0.12 & H, C \\
2002cr   & NGC5468          &   3.45 &   0.94$\pm$0.02 &  14.20$\pm$0.02 & C \\
2002de   & NGC6104          &   3.92 &   1.06$\pm$0.07 &  16.68$\pm$0.02 & H, C \\
2002dj   & NGC5018          &   3.45 &   0.94$\pm$0.01 &  13.93$\pm$0.04 & C \\
2002dp   & NGC7678          &   3.54 &   0.96$\pm$0.04 &  14.61$\pm$0.03 & C \\
2002er   & UGC10743         &   3.43 &   0.88$\pm$0.01 &  14.26$\pm$0.06 & C \\
2002es   & UGC2708          &   3.73 &   1.01$\pm$0.05 &  16.16$\pm$0.10 & H \\
2002fk   & NGC1309          &   3.35 &   0.99$\pm$0.03 &  13.13$\pm$0.03 & C \\
2002ha   & NGC6962          &   3.63 &   0.87$\pm$0.03 &  14.71$\pm$0.05 & H, C \\
2002he   & UGC4322          &   3.87 &   0.80$\pm$0.02 &  16.23$\pm$0.04 & H, C \\
2002hu   & M+06-06-12       &   4.04 &   1.04$\pm$0.02 &  16.63$\pm$0.02 & H, C \\
2002hw   & UGC52            &   3.72 &   0.76$\pm$0.03 &  16.65$\pm$0.05 & L, H \\
2002jy   & NGC477           &   3.77 &   1.12$\pm$0.04 &  15.74$\pm$0.03 & H, C \\
2003U    & NGC6365A         &   3.93 &   0.79$\pm$0.02 &  16.51$\pm$0.03 & L, H, C \\
2003W    & UGC5234          &   3.78 &   0.99$\pm$0.02 &  15.88$\pm$0.02 & H, C \\
2003cg   & NGC3169          &   3.09 &   0.95$\pm$0.01 &  15.79$\pm$0.02 & R, C \\
2003du   & UGC9391          &   3.28 &   1.01$\pm$0.01 &  13.47$\pm$0.01 & C \\
2003fa   & M+07-36-33       &   4.07 &   1.15$\pm$0.01 &  16.71$\pm$0.02 & H, C \\
2003hu   & A191131+7753     &   4.35 &   1.16$\pm$0.06 &  18.46$\pm$0.14 & H, C \\
2003hx   & NGC2076          &   3.33 &   0.84$\pm$0.06 &  14.86$\pm$0.05 & C \\
2003ic   & M-02-02-86       &   4.22 &   0.75$\pm$0.05 &  17.66$\pm$0.08 & L, H \\
2003kc   & M+05-23-37       &   4.00 &   0.84$\pm$0.04 &  17.14$\pm$0.05 & H \\
2003kf   & M-02-16-02       &   3.35 &   1.04$\pm$0.03 &  13.27$\pm$0.13 & C \\
2004L    & M+03-27-38       &   3.99 &   0.93$\pm$0.04 &  17.30$\pm$0.05 & H, C \\
2004as   & A112539+2249     &   3.97 &   1.06$\pm$0.04 &  16.96$\pm$0.02 & H, C \\
2004eo   & NGC6928          &   3.67 &   0.87$\pm$0.00 &  15.08$\pm$0.05 & H, C \\
2004fu   & NGC6949          &   3.44 &   0.87$\pm$0.01 &  14.25$\pm$0.16 & C \\
2005am   & NGC2811          &   3.40 &   0.70$\pm$0.05 &  13.66$\pm$0.03 & L, C \\
2005eq   & M-01-09-06       &   3.95 &   1.18$\pm$0.01 &  16.30$\pm$0.03 & H, C \\
2005hc   & M+00-06-03       &   4.14 &   1.02$\pm$0.03 &  17.36$\pm$0.02 & H, C \\
2005hk   & UGC272           &   3.59 &   0.88$\pm$0.00 &  15.95$\pm$0.01 &  \\
2005iq   & M-03-01-08       &   4.01 &   0.87$\pm$0.02 &  16.82$\pm$0.02 & H, C \\
2005ir   & A011643+0047     &   4.36 &   1.44$\pm$0.11 &  18.42$\pm$0.03 & H, C \\
2005kc   & NGC7311          &   3.65 &   0.93$\pm$0.02 &  15.61$\pm$0.06 & H, C \\
2005ke   & NGC1371          &   3.16 &   0.64$\pm$0.04 &  14.80$\pm$0.03 & L \\
2005ki   & NGC3332          &   3.76 &   0.80$\pm$0.01 &  15.55$\pm$0.03 & H, C \\
2005ls   & M+07-07-01       &   3.80 &   1.13$\pm$0.03 &  16.25$\pm$0.04 & H, C \\
2005mc   & UGC4414          &   3.88 &   0.65$\pm$0.04 &  17.23$\pm$0.02 & L, H \\
2005ms   & UGC4614          &   3.88 &   1.06$\pm$0.02 &  16.16$\pm$0.02 & H, C \\
2005mz   & NGC1275          &   3.72 &   0.61$\pm$0.02 &  16.42$\pm$0.07 & L, H \\
2006N    & M+11-08-12       &   3.63 &   0.76$\pm$0.02 &  15.08$\pm$0.04 & L, H, C \\
2006S    & UGC7934          &   3.98 &   1.12$\pm$0.01 &  16.86$\pm$0.01 & H, C \\
2006X    & NGC4321          &   3.20 &   0.96$\pm$0.01 &  15.22$\pm$0.01 & R, C \\
2006ac   & NGC4619          &   3.83 &   0.87$\pm$0.02 &  16.18$\pm$0.02 & H, C \\
2006ak   & A110932+2837     &   4.05 &   0.84$\pm$0.04 &  17.24$\pm$0.10 & H, C \\
2006al   & A103929+0511     &   4.31 &   0.78$\pm$0.04 &  18.44$\pm$0.05 & L, H, C \\
2006ar   & M+11-13-36       &   3.83 &   0.92$\pm$0.03 &  16.48$\pm$0.01 & H, C \\
2006ax   & NGC3663          &   3.70 &   1.00$\pm$0.01 &  15.04$\pm$0.02 & H, C \\
2006az   & NGC4172          &   3.97 &   0.86$\pm$0.01 &  16.49$\pm$0.01 & H, C \\
2006bk   & M+06-33-20       &   4.17 &   1.10$\pm$0.03 &  17.00$\pm$0.06 & H \\
2006bq   & NGC6685          &   3.82 &   0.84$\pm$0.02 &  16.15$\pm$0.04 & H, C \\
2006br   & NGC5185          &   3.87 &   0.81$\pm$0.04 &  18.95$\pm$0.03 & R, H, C \\
2006bt   & M+03-41-04       &   3.98 &   1.01$\pm$0.02 &  16.95$\pm$0.02 & H, C \\
2006bw   & A143356+0347     &   4.00 &   0.72$\pm$0.03 &  17.56$\pm$0.04 & L, H, C \\
2006cc   & UGC10244         &   3.99 &   1.03$\pm$0.01 &  17.81$\pm$0.01 & H, C \\
2006cj   & A125924+2820     &   4.31 &   1.25$\pm$0.11 &  18.14$\pm$0.03 & H, C \\
2006cm   & UGC11723         &   3.69 &   1.05$\pm$0.04 &  17.94$\pm$0.03 & R, H, C \\
2006cp   & UGC7357          &   3.82 &   1.07$\pm$0.02 &  15.99$\pm$0.02 & H, C \\
2006ef   & NGC809           &   3.73 &   0.84$\pm$0.04 &  15.16$\pm$0.17 & H \\
2006ej   & NGC191A          &   3.79 &   0.81$\pm$0.04 &  15.79$\pm$0.03 & H, C \\
2006en   & M+05-54-41       &   3.98 &   1.00$\pm$0.04 &  16.75$\pm$0.05 & H, C \\
2006gj   & UGC2650          &   3.93 &   0.65$\pm$0.07 &  17.61$\pm$0.05 & L, H, C \\
2006gz   & IC1277           &   3.85 &   1.26$\pm$0.01 &  15.84$\pm$0.03 & H \\
2006hb   & M-04-12-34       &   3.66 &   0.67$\pm$0.06 &  15.48$\pm$0.05 & L, H, C \\
2006kf   & UGC2829          &   3.80 &   0.71$\pm$0.03 &  15.83$\pm$0.10 & L, H, C \\
2006le   & UGC3218          &   3.72 &   1.11$\pm$0.01 &  14.80$\pm$0.17 & H, C \\
2006mo   & M+06-02-17       &   4.05 &   0.75$\pm$0.03 &  17.44$\pm$0.03 & L, H, C \\
2006nz   & A005629-0113     &   4.06 &   0.60$\pm$0.10 &  18.08$\pm$0.04 & L, H \\
2006ob   & UGC1333          &   4.25 &   0.71$\pm$0.01 &  18.23$\pm$0.02 & L, H, C \\
2006on   & A215558-0104     &   4.32 &   0.99$\pm$0.09 &  18.41$\pm$0.07 & H, C \\
2006os   & UGC2384          &   3.99 &   0.91$\pm$0.03 &  17.61$\pm$0.06 & H \\
2006sr   & UGC14            &   3.86 &   0.84$\pm$0.02 &  16.13$\pm$0.04 & H, C \\
2006te   & A081144+4133     &   3.98 &   1.03$\pm$0.05 &  16.51$\pm$0.04 & H, C \\
2007F    & UGC8162          &   3.85 &   1.07$\pm$0.01 &  15.90$\pm$0.01 & H, C \\
2007O    & UGC9612          &   4.03 &   0.93$\pm$0.04 &  16.78$\pm$0.05 & H, C \\
2007R    & UGC4008          &   3.96 &   0.82$\pm$0.02 &  16.65$\pm$0.05 & H \\
2007S    & UGC5378          &   3.62 &   1.13$\pm$0.01 &  15.82$\pm$0.01 & H, C \\
2007ae   & UGC10704         &   4.29 &   1.15$\pm$0.03 &  17.78$\pm$0.04 & H, C \\
2007af   & NGC5584          &   3.20 &   0.96$\pm$0.01 &  13.16$\pm$0.02 & C \\
2007ap   & M+03-41-03       &   3.67 &   0.54$\pm$0.07 &  15.86$\pm$0.03 & L, H, C \\
2007au   & UGC3725          &   3.79 &   0.66$\pm$0.03 &  16.51$\pm$0.03 & L, H, C \\
2007bc   & UGC6332          &   3.80 &   0.84$\pm$0.03 &  15.89$\pm$0.02 & H, C \\
2007bd   & UGC4455          &   3.97 &   0.82$\pm$0.01 &  16.57$\pm$0.02 & H, C \\
2007bm   & NGC3672          &   3.27 &   0.92$\pm$0.01 &  14.48$\pm$0.02 & C \\
2007bz   & IC3918           &   3.81 &   1.17$\pm$0.03 &  16.67$\pm$0.03 & H, C \\
2007cg   & E508-G75         &   4.00 &   0.82$\pm$0.05 &  18.28$\pm$0.07 & H, C \\
2007ci   & NGC3873          &   3.74 &   0.75$\pm$0.01 &  15.92$\pm$0.02 & L, H, C \\
2007sr   & NGC4038          &   3.21 &   0.99$\pm$0.02 &  12.76$\pm$0.04 & C \\
2008af   & UGC9640          &   4.00 &   0.85$\pm$0.03 &  16.78$\pm$0.08 & H, C \\
2008bf   & NGC4055          &   3.86 &   1.04$\pm$0.02 &  15.73$\pm$0.02 & H, C
\tablenotetext{1}{stretch errors for SNe with $s < 0.7$ have been multiplied by 3.0}
\tablenotetext{2}{{L} - low stretch: $s<0.80$, {R} - red: $\mathcal{C}>0.7$, H - in Hubble flow: $z>0.0133$, C - eligible for cosmology fitting}
\end{longtable}

\clearpage

\begin{longtable}{lrrrrrrrrrrc}
\tablecaption{Host Properties\label{tab_host_sed}}
\tablewidth{0pt}
\tablehead{
 & &
\colhead{Age$-$} & \colhead{$<$Age$>_L$} & \colhead{Age$+$} &
\colhead{M*$-$} & \colhead{$<$M*$>$} & \colhead{M*$+$} &
\colhead{sSFR$-$} & \colhead{$<$sSFR$>$} & \colhead{sSFR$+$} &
\colhead{E(B-V)$_{H}$} \\
\colhead{HOST} & \colhead{T\tablenotemark{1}} &
\multicolumn{3}{c}{($\log$ yr)} &
\multicolumn{3}{c}{($\log$ M$_{\odot}$)} &
\multicolumn{3}{c}{($\log$ yr$^{-1}$)} &
\colhead{(Mag)}
}
A005629-0113     & \ldots &   9.81 &  10.07 &  10.09 &  10.48 &  10.62 &  10.66 &  -12.00 &  -12.00 &  -12.48 &  0.00 \\
A011643+0047     & \ldots &   8.45 &   8.71 &   8.93 &  10.06 &  10.15 &  10.20 &   -9.50 &   -9.31 &   -9.06 &  0.20 \\
A081144+4133     & \ldots &   8.83 &   9.13 &   9.39 &  10.26 &  10.31 &  10.43 &  -10.32 &   -9.84 &   -9.41 &  0.10 \\
A103929+0511     & \ldots &   9.51 &   9.51 &  10.05 &  10.23 &  10.26 &  10.55 &  -12.00 &  -10.76 &  -10.71 &  0.05 \\
A110932+2837     & \ldots &   8.86 &   8.97 &   9.56 &  10.51 &  10.52 &  10.75 &  -10.77 &  -10.09 &   -9.75 &  0.30 \\
A112539+2249     & \ldots &   8.15 &   8.62 &   9.05 &   9.15 &   9.28 &   9.36 &   -9.58 &   -9.23 &   -8.79 &  0.05 \\
A125924+2820     & \ldots &   8.15 &   9.03 &   9.44 &  10.30 &  10.42 &  10.62 &  -10.69 &  -10.13 &   -8.35 &  0.20 \\
A143356+0347     & \ldots &   9.51 &   9.62 &  10.09 &   9.95 &  10.03 &  10.27 &  -12.00 &  -11.03 &  -10.69 &  0.05 \\
A191131+7753     & \ldots &   8.47 &   8.66 &   8.95 &  10.86 &  10.90 &  10.98 &   -9.52 &   -9.27 &   -9.07 &  0.25 \\
A215558-0104     & \ldots &   9.51 &   9.55 &   9.65 &  10.25 &  10.30 &  10.38 &  -12.00 &  -11.62 &  -10.73 &  0.00 \\
CGCG180-22       & \ldots &   9.04 &   9.09 &   9.28 &  10.10 &  10.13 &  10.21 &  -10.41 &  -10.14 &  -10.06 &  0.15 \\
CGCG266-031      &   3.0 &   9.15 &   9.37 &   9.61 &  10.55 &  10.62 &  10.74 &  -10.83 &  -10.49 &  -10.17 &  0.05 \\
E291-G11         &   1.0 &   8.90 &   9.73 &  10.03 &  11.16 &  11.81 &  12.27 &  -11.56 &  -10.85 &   -9.27 &  0.30 \\
E300-G09         &   5.0 &   8.50 &   8.91 &   9.59 &   9.19 &   9.72 &  10.42 &  -10.56 &   -9.62 &   -8.50 &  0.10 \\
E352-G57         &  -1.5 &   9.16 &   9.99 &  10.11 &  10.93 &  12.13 &  12.27 &  -12.00 &  -11.18 &   -9.49 &  0.60 \\
E383-G32         &   4.5 &   8.08 &   8.35 &   8.71 &   9.70 &  10.04 &  10.27 &   -9.35 &   -9.00 &   -8.49 &  0.25 \\
E445-G66         &   1.9 &   7.00 &   7.76 &   9.15 &  10.13 &  10.51 &  11.07 &  -10.37 &   -8.63 &   -8.14 &  0.55 \\
E508-G67         &   5.0 &   8.81 &   8.90 &   9.67 &   9.92 &  10.02 &  11.06 &  -11.11 &   -9.88 &   -8.95 &  0.00 \\
E508-G75         &   3.9 &   7.83 &   9.36 &  10.03 &  10.13 &  10.76 &  11.35 &  -11.57 &  -10.31 &   -8.14 &  0.15 \\
E576-G40         &   7.0 &   7.00 &   7.30 &   8.00 &   8.59 &   8.71 &   8.90 &   -8.97 &   -8.66 &   -8.33 &  0.45 \\
IC1277           &   6.0 &   7.00 &   7.26 &   8.21 &  10.29 &  10.44 &  10.88 &   -9.26 &   -8.66 &   -8.15 &  0.55 \\
IC3690           &   4.0 &   8.83 &   9.16 &   9.67 &  10.24 &  10.34 &  10.48 &  -10.86 &   -9.98 &   -9.35 &  0.15 \\
IC3918           &   4.4 &   8.01 &   8.28 &   8.59 &   9.27 &   9.38 &   9.45 &   -9.19 &   -8.94 &   -8.62 &  0.15 \\
IC4232           &   3.8 &   7.71 &   8.73 &   9.48 &  10.34 &  11.04 &  11.51 &  -11.03 &   -9.32 &   -8.15 &  0.40 \\
IC4423           &   4.1 &   8.88 &   9.18 &   9.72 &  10.49 &  10.62 &  10.77 &  -10.89 &  -10.02 &   -9.49 &  0.15 \\
IC4919           &   7.9 &   7.28 &   7.30 &   7.51 &   9.04 &   9.07 &   9.10 &   -8.72 &   -8.66 &   -8.58 &  0.25 \\
IC5179           &   4.0 &   8.68 &   8.86 &   8.97 &  10.21 &  10.72 &  10.83 &  -10.50 &   -9.41 &   -8.86 &  0.35 \\
M+00-06-03       &  -2.0 &   8.88 &   9.13 &   9.68 &  10.45 &  10.54 &  10.77 &  -10.85 &  -10.15 &   -9.42 &  0.15 \\
M+01-57-14       &   6.0 &   8.15 &   9.94 &  10.05 &  10.57 &  11.32 &  11.79 &  -11.79 &  -11.10 &   -8.16 &  0.20 \\
M+03-27-38       &   5.0 &   7.57 &   9.10 &   9.29 &  10.20 &  10.35 &  10.56 &  -10.49 &   -9.90 &   -8.55 &  0.05 \\
M+03-41-03       &  -1.0 &   8.78 &   8.92 &  10.09 &  10.29 &  10.35 &  10.69 &  -12.00 &   -9.96 &   -9.49 &  0.30 \\
M+03-41-04       &   0.0 &   8.50 &   8.84 &  10.10 &  11.02 &  11.09 &  11.35 &  -12.00 &   -9.93 &   -8.54 &  0.40 \\
M+05-23-37       &   3.0 &   8.54 &   8.82 &   9.10 &  10.53 &  10.67 &  10.76 &   -9.99 &   -9.38 &   -9.07 &  0.25 \\
M+05-54-41       &   5.0 &   7.45 &   8.24 &   8.97 &  10.23 &  10.69 &  11.23 &   -9.78 &   -8.92 &   -8.25 &  0.45 \\
M+06-02-17       &   4.2 &   9.51 &   9.51 &   9.86 &  10.80 &  10.82 &  10.99 &  -12.00 &  -10.75 &  -10.71 &  0.05 \\
M+06-06-12       &   4.8 &   7.00 &   9.09 &   9.70 &   8.83 &  10.27 &  10.93 &  -10.95 &   -9.68 &   -8.00 &  0.15 \\
M+06-33-20       &  -3.1 &   8.76 &   9.56 &  10.09 &  11.49 &  11.59 &  11.83 &  -12.00 &  -10.82 &   -9.46 &  0.10 \\
M+07-07-01       &   4.2 &   7.28 &   7.81 &   8.53 &   9.66 &   9.86 &  10.04 &   -9.16 &   -8.63 &   -8.35 &  0.30 \\
M+07-36-33       & \ldots &   8.32 &   8.92 &   9.59 &  10.03 &  10.81 &  11.42 &  -11.03 &   -9.57 &   -8.44 &  0.20 \\
M+08-25-47       & \ldots &   8.83 &   9.16 &   9.39 &   9.96 &   9.99 &  10.16 &  -10.37 &   -9.98 &   -9.38 &  0.05 \\
M+11-08-12       &  -2.4 &   9.51 &  10.10 &  10.11 &  10.59 &  10.82 &  10.88 &  -12.00 &  -12.00 &  -10.71 &  0.00 \\
M+11-13-36       &   3.3 &   8.58 &   9.15 &   9.24 &   9.67 &   9.72 &  10.18 &  -10.69 &  -10.18 &   -8.81 &  0.00 \\
M-01-04-44       &   6.0 &   7.54 &   7.98 &   8.15 &   8.91 &   8.93 &   9.20 &   -8.91 &   -8.63 &   -8.41 &  0.25 \\
M-01-09-06       &   6.0 &   8.84 &   9.08 &   9.31 &  10.51 &  10.58 &  10.77 &  -10.53 &  -10.14 &   -9.36 &  0.05 \\
M-02-02-86       &  -2.0 &   9.51 &   9.51 &  10.09 &  11.67 &  11.70 &  11.97 &  -12.00 &  -10.76 &  -10.69 &  0.10 \\
M-02-16-02       &   3.0 &   7.00 &   9.03 &  10.05 &   9.29 &   9.58 &  10.50 &  -11.39 &  -10.13 &   -8.07 &  0.05 \\
M-03-01-08       &   2.5 &   8.79 &   9.05 &   9.70 &  10.17 &  10.34 &  10.95 &  -10.63 &   -9.81 &   -9.20 &  0.05 \\
M-04-12-34       &  -2.6 &   8.43 &   8.79 &   9.61 &  10.64 &  10.95 &  11.14 &  -11.34 &   -9.81 &   -8.16 &  0.55 \\
M-05-28-01       &   3.7 &   7.67 &   8.28 &   9.15 &  10.48 &  10.98 &  11.41 &  -10.32 &   -8.94 &   -8.17 &  0.45 \\
NGC105           &   1.5 &   7.04 &   9.11 &   9.59 &   9.47 &  10.87 &  11.42 &  -11.04 &   -9.81 &   -8.03 &  0.15 \\
NGC1275          &  -1.6 &   8.51 &   9.05 &   9.36 &  11.15 &  11.24 &  11.48 &  -10.49 &   -9.81 &   -9.03 &  0.15 \\
NGC1309          &   4.0 &   7.82 &   8.47 &   8.84 &   9.78 &   9.94 &  10.64 &  -10.27 &   -9.08 &   -8.00 &  0.25 \\
NGC1316          &  -1.9 &  10.07 &  10.11 &  10.11 &  11.75 &  11.79 &  11.80 &  -12.00 &  -12.00 &  -13.75 &  0.00 \\
NGC1371          &   1.0 &   9.51 &   9.70 &   9.77 &  10.79 &  10.90 &  11.02 &  -10.60 &  -10.46 &  -10.15 &  0.15 \\
NGC1380          &  -1.9 &  10.05 &  10.10 &  10.10 &  11.27 &  11.33 &  11.36 &  -12.00 &  -12.00 &  -13.27 &  0.05 \\
NGC1448          &   5.9 &   8.95 &   9.37 &   9.40 &  10.22 &  10.69 &  10.73 &  -10.19 &   -9.90 &   -9.40 &  0.20 \\
NGC1699          &   3.0 &   8.71 &   9.59 &   9.62 &  10.12 &  10.37 &  10.41 &  -10.46 &  -10.22 &   -9.24 &  0.00 \\
NGC191A          &  -2.0 &   9.51 &  10.10 &  10.11 &  10.76 &  11.00 &  11.02 &  -12.00 &  -12.00 &  -10.70 &  0.00 \\
NGC2076          &  -0.6 &   9.18 &   9.67 &   9.77 &   9.62 &  10.02 &  10.14 &  -11.12 &  -10.41 &   -9.62 &  0.15 \\
NGC2258          &  -2.0 &   8.43 &   8.91 &   9.48 &  11.43 &  11.52 &  12.33 &  -11.01 &  -10.03 &   -8.40 &  0.50 \\
NGC2441          &   3.1 &   8.71 &   9.12 &   9.40 &  10.31 &  10.66 &  10.87 &  -10.04 &   -9.70 &   -9.28 &  0.25 \\
NGC252           &  -1.0 &   8.94 &   9.43 &   9.90 &  10.87 &  11.25 &  11.90 &  -11.44 &  -10.56 &   -9.43 &  0.25 \\
NGC2595          &   4.5 &   8.30 &   9.19 &   9.38 &  10.62 &  10.72 &  10.96 &  -10.55 &  -10.23 &   -8.84 &  0.00 \\
NGC2623          &  PEC  &   8.21 &   8.93 &   9.34 &  10.27 &  10.38 &  10.52 &  -10.50 &  -10.05 &   -8.43 &  0.10 \\
NGC2811          &   1.0 &   8.90 &   9.26 &   9.56 &  11.28 &  11.51 &  11.71 &  -10.81 &  -10.37 &   -9.85 &  0.45 \\
NGC2935          &   3.2 &   7.58 &   8.25 &  10.05 &  10.68 &  10.94 &  11.31 &  -12.00 &   -8.61 &   -8.23 &  0.55 \\
NGC2962          &  -1.0 &   8.63 &   8.93 &  10.11 &  10.10 &  10.20 &  10.50 &  -12.00 &  -10.05 &   -8.95 &  0.35 \\
NGC3021          &   4.3 &   8.71 &   8.90 &   9.19 &   9.76 &   9.87 &   9.97 &   -9.88 &   -9.45 &   -9.23 &  0.30 \\
NGC3147          &   3.9 &   8.98 &   9.34 &   9.63 &  10.75 &  11.31 &  11.66 &  -11.02 &  -10.15 &   -9.41 &  0.15 \\
NGC3169          &   1.2 &   9.04 &   9.19 &   9.43 &  10.64 &  10.70 &  10.79 &  -10.59 &  -10.23 &  -10.07 &  0.15 \\
NGC3190          &   1.0 &   9.51 &  10.11 &  10.11 &  10.55 &  10.83 &  10.89 &  -12.00 &  -12.00 &  -10.62 &  0.00 \\
NGC3327          &   3.0 &   8.79 &   9.18 &   9.75 &  10.66 &  10.77 &  10.96 &  -10.92 &  -10.02 &   -9.36 &  0.15 \\
NGC3332          &  -3.0 &   8.72 &   8.96 &  10.11 &  11.11 &  11.15 &  11.51 &  -12.00 &  -10.07 &   -9.29 &  0.30 \\
NGC3368          &   2.0 &   8.80 &   9.22 &   9.83 &  11.09 &  11.26 &  11.54 &  -11.01 &  -10.30 &   -9.59 &  0.25 \\
NGC3370          &   5.3 &   8.00 &   8.55 &   9.08 &   9.55 &   9.69 &   9.83 &   -9.97 &   -9.15 &   -8.63 &  0.25 \\
NGC3663          &   3.5 &   8.53 &   9.08 &   9.44 &  10.04 &  10.81 &  11.21 &  -11.00 &   -9.60 &   -8.54 &  0.30 \\
NGC3672          &   5.0 &   7.92 &   8.65 &   8.97 &  10.04 &  10.23 &  10.34 &   -9.52 &   -9.23 &   -8.58 &  0.35 \\
NGC382           &  -5.0 &   9.13 &  10.05 &  10.05 &  10.17 &  11.54 &  11.56 &  -11.80 &  -11.24 &   -9.63 &  0.45 \\
NGC3873          &  -5.0 &   9.51 &  10.00 &  10.11 &  10.93 &  11.13 &  11.23 &  -12.00 &  -12.00 &  -10.68 &  0.00 \\
NGC3982          &   3.0 &   8.69 &   9.25 &   9.40 &   9.86 &  10.02 &  10.10 &  -10.12 &   -9.75 &   -9.20 &  0.10 \\
NGC3987          &   3.0 &   9.08 &   9.30 &   9.77 &  10.69 &  10.78 &  10.97 &  -10.93 &  -10.43 &  -10.08 &  0.25 \\
NGC4038          &   8.8 &   7.43 &   7.79 &   8.15 &   9.85 &  10.05 &  10.17 &   -8.95 &   -8.64 &   -8.38 &  0.30 \\
NGC4055          &  -5.0 &   9.51 &  10.11 &  10.11 &  11.11 &  11.39 &  11.42 &  -12.00 &  -12.00 &  -10.66 &  0.00 \\
NGC4172          &   2.7 &   9.51 &   9.51 &  10.11 &  11.26 &  11.28 &  11.54 &  -12.00 &  -10.76 &  -10.72 &  0.10 \\
NGC4321          &   4.4 &   8.57 &   9.17 &   9.26 &  10.74 &  10.81 &  11.09 &  -10.57 &  -10.22 &   -9.01 &  0.00 \\
NGC4493          &  -4.0 &   8.90 &   9.61 &  10.11 &  10.93 &  11.04 &  11.23 &  -12.00 &  -10.88 &   -9.87 &  0.10 \\
NGC4495          &   1.9 &   8.92 &   9.16 &   9.19 &  10.39 &  10.50 &  10.53 &   -9.72 &   -9.68 &   -9.54 &  0.30 \\
NGC4501          &   3.0 &   8.36 &   8.79 &   9.24 &  10.91 &  11.03 &  11.23 &  -10.53 &   -9.45 &   -8.86 &  0.35 \\
NGC4527          &   3.8 &   8.62 &   9.30 &   9.56 &  10.60 &  10.78 &  10.91 &  -10.79 &  -10.43 &   -9.03 &  0.20 \\
NGC4536          &   4.5 &   9.00 &   9.18 &   9.36 &  10.42 &  10.47 &  10.68 &  -10.45 &  -10.02 &   -9.38 &  0.05 \\
NGC4619          &   3.1 &   8.44 &   8.88 &   9.08 &  10.87 &  10.92 &  11.11 &   -9.97 &   -9.56 &   -9.01 &  0.15 \\
NGC4639          &   3.8 &   9.16 &   9.32 &   9.46 &  10.07 &  10.14 &  10.24 &  -10.42 &  -10.12 &   -9.63 &  0.05 \\
NGC4675          &   3.0 &   8.56 &   9.62 &   9.77 &  10.14 &  10.42 &  10.43 &  -10.93 &  -10.26 &   -9.10 &  0.10 \\
NGC4680          &  PEC  &   8.18 &   8.76 &   9.05 &  10.11 &  10.24 &  10.33 &   -9.92 &   -9.32 &   -8.81 &  0.30 \\
NGC4704          &   3.5 &   8.35 &   8.95 &   9.25 &  10.55 &  10.59 &  10.82 &  -10.31 &   -9.67 &   -8.95 &  0.15 \\
NGC477           &   5.0 &   8.31 &   8.88 &   9.19 &  10.32 &  10.46 &  10.57 &   -9.89 &   -9.43 &   -8.95 &  0.20 \\
NGC5005          &   4.0 &   8.79 &   8.96 &   9.00 &  10.90 &  10.93 &  11.01 &  -10.19 &  -10.07 &   -9.61 &  0.30 \\
NGC5018          &  -5.0 &   9.51 &   9.51 &   9.61 &  11.21 &  11.34 &  11.37 &  -11.24 &  -10.76 &  -10.60 &  0.10 \\
NGC5061          &  -5.0 &   9.51 &   9.61 &   9.61 &  10.78 &  10.85 &  10.87 &  -10.92 &  -10.88 &  -10.67 &  0.05 \\
NGC5185          &   3.0 &   9.18 &   9.37 &   9.61 &  10.89 &  10.94 &  11.06 &  -10.83 &  -10.49 &  -10.20 &  0.10 \\
NGC523           &   4.2 &   8.95 &   9.32 &   9.51 &  10.23 &  10.38 &  10.53 &  -10.71 &  -10.45 &   -9.56 &  0.00 \\
NGC524           &  -1.0 &  10.05 &  10.10 &  10.10 &  11.82 &  11.89 &  11.91 &  -12.00 &  -12.00 &  -13.82 &  0.05 \\
NGC5468          &   6.0 &   7.18 &   7.43 &   7.87 &   9.39 &   9.48 &   9.77 &   -9.04 &   -8.65 &   -8.25 &  0.20 \\
NGC5584          &   6.0 &   8.27 &   8.71 &   8.92 &   9.69 &   9.82 &   9.87 &   -9.49 &   -9.31 &   -8.94 &  0.15 \\
NGC6038          &   5.0 &   8.91 &   9.18 &   9.34 &  10.94 &  10.99 &  11.03 &  -10.29 &  -10.02 &   -9.60 &  0.10 \\
NGC6063          &   6.0 &   8.56 &   8.95 &   9.21 &   9.85 &   9.89 &  10.11 &  -10.11 &   -9.67 &   -9.10 &  0.10 \\
NGC6104          & \ldots &   8.92 &   9.16 &   9.19 &  10.71 &  10.83 &  10.86 &   -9.90 &   -9.68 &   -9.51 &  0.20 \\
NGC632           &  -1.5 &   7.89 &   9.24 &   9.35 &   9.83 &  10.00 &  10.13 &  -10.57 &  -10.34 &   -8.43 &  0.00 \\
NGC6365A         &   5.9 &   8.43 &   9.40 &   9.59 &  10.44 &  10.74 &  10.87 &  -10.26 &   -9.93 &   -9.02 &  0.05 \\
NGC6411          &  -5.0 &   9.51 &   9.70 &  10.11 &  10.79 &  10.91 &  11.13 &  -12.00 &  -12.00 &  -10.67 &  0.05 \\
NGC6627          &   3.0 &   8.88 &   9.42 &   9.78 &  10.68 &  10.85 &  11.08 &  -12.00 &  -10.56 &   -9.44 &  0.05 \\
NGC6685          &  -3.0 &   8.72 &   8.96 &  10.11 &  10.73 &  10.76 &  11.12 &  -12.00 &  -10.07 &   -9.31 &  0.30 \\
NGC673           &   5.0 &   7.81 &   8.13 &   8.48 &  10.24 &  10.37 &  10.78 &   -9.39 &   -8.84 &   -8.32 &  0.30 \\
NGC6916          &   4.0 &   8.69 &   9.59 &   9.62 &  10.43 &  10.68 &  10.72 &  -10.46 &  -10.22 &   -9.22 &  0.00 \\
NGC6928          &   2.0 &   8.15 &   8.83 &   8.97 &  11.05 &  11.17 &  11.40 &  -10.37 &   -9.93 &   -8.26 &  0.40 \\
NGC6949          &   5.0 &   7.00 &   8.17 &   9.37 &   9.09 &   9.64 &  10.42 &  -10.02 &   -8.86 &   -8.04 &  0.10 \\
NGC6951          &   4.0 &   7.88 &   8.63 &   9.61 &  10.84 &  10.96 &  11.32 &  -10.94 &   -9.05 &   -8.18 &  0.35 \\
NGC6962          &   2.0 &   8.83 &   9.30 &   9.56 &  10.96 &  11.09 &  11.23 &  -10.79 &  -10.43 &   -9.67 &  0.10 \\
NGC7311          &   2.0 &   8.94 &   9.32 &   9.56 &  10.84 &  10.97 &  11.09 &  -10.78 &  -10.45 &   -9.94 &  0.10 \\
NGC7541          &   4.0 &   7.18 &   8.48 &   9.70 &  10.29 &  10.66 &  10.98 &  -10.82 &   -9.11 &   -8.22 &  0.35 \\
NGC7678          &   5.0 &   8.67 &   8.74 &   9.05 &  10.04 &  10.40 &  10.75 &  -10.37 &   -9.38 &   -8.81 &  0.15 \\
NGC7780          &   2.0 &   8.94 &   9.30 &   9.61 &  10.22 &  10.35 &  10.48 &  -10.87 &  -10.43 &   -9.93 &  0.15 \\
NGC809           &  -1.0 &   9.51 &   9.51 &   9.61 &  10.66 &  10.70 &  10.78 &  -10.99 &  -10.75 &  -10.69 &  0.05 \\
NGC976           &   5.0 &   8.66 &   9.01 &   9.46 &  10.72 &  10.78 &  10.99 &  -10.60 &   -9.77 &   -9.16 &  0.15 \\
PGC0059076       & \ldots &   8.64 &   8.98 &   9.13 &   9.80 &   9.84 &  10.00 &  -10.08 &   -9.72 &   -9.14 &  0.05 \\
UGC10030         &   3.0 &   8.88 &   9.27 &   9.77 &  10.89 &  10.97 &  11.19 &  -10.94 &  -10.16 &   -9.47 &  0.10 \\
UGC10244         &   3.7 &   8.62 &   9.07 &   9.76 &  10.41 &  10.46 &  10.65 &  -10.88 &   -9.86 &   -9.15 &  0.25 \\
UGC10483         &   3.8 &   8.41 &   8.74 &   9.22 &  10.55 &  10.68 &  10.82 &  -10.40 &   -9.38 &   -8.87 &  0.35 \\
UGC10704         &   3.5 &   8.19 &   9.30 &   9.61 &  11.25 &  11.44 &  11.61 &  -10.84 &  -10.43 &   -8.44 &  0.10 \\
UGC1071          & \ldots &   8.21 &   9.51 &   9.62 &  10.29 &  10.35 &  11.90 &  -12.00 &  -10.76 &   -8.00 &  0.00 \\
UGC10743         &   1.0 &   7.88 &   8.51 &   9.77 &   9.93 &  10.56 &  11.18 &  -11.42 &   -9.13 &   -8.08 &  0.70 \\
UGC1087          &   5.0 &   8.69 &   9.26 &   9.48 &  10.04 &  10.20 &  10.29 &  -10.14 &   -9.77 &   -9.25 &  0.10 \\
UGC11149         & \ldots &   7.98 &   9.61 &  10.09 &  11.17 &  11.72 &  12.59 &  -12.00 &  -12.00 &   -8.00 &  0.05 \\
UGC11723         &   3.0 &   9.01 &   9.50 &   9.87 &  10.22 &  10.41 &  10.58 &  -11.05 &  -10.65 &   -9.62 &  0.15 \\
UGC1333          &   3.0 &   8.99 &   9.05 &   9.24 &  11.20 &  11.25 &  11.34 &  -10.34 &  -10.13 &  -10.01 &  0.20 \\
UGC14            &   5.5 &   7.86 &   8.56 &   9.25 &  10.52 &  10.70 &  10.87 &  -10.20 &   -9.16 &   -8.59 &  0.35 \\
UGC2384          &   3.9 &   7.40 &   8.54 &   9.74 &  10.66 &  11.59 &  12.43 &  -11.42 &   -9.15 &   -8.00 &  0.70 \\
UGC2650          &   2.0 &   8.50 &   9.98 &  10.11 &  10.66 &  11.96 &  12.33 &  -12.00 &  -11.11 &   -8.07 &  0.45 \\
UGC2708          &  -2.0 &   8.95 &   9.86 &  10.00 &  10.81 &  11.08 &  11.19 &  -12.00 &  -12.00 &   -9.98 &  0.00 \\
UGC272           &   6.5 &   8.00 &   8.46 &   8.99 &   9.41 &   9.54 &   9.68 &   -9.69 &   -9.07 &   -8.68 &  0.20 \\
UGC2829          &  -2.0 &   8.99 &   9.86 &  10.01 &  10.73 &  10.97 &  11.06 &  -12.00 &  -12.00 &  -10.09 &  0.00 \\
UGC3151          &   5.0 &   8.70 &   9.21 &   9.77 &  10.53 &  10.65 &  10.92 &  -12.00 &  -10.06 &   -9.16 &  0.10 \\
UGC3218          &   3.0 &   7.00 &   7.34 &   7.75 &  10.09 &  10.19 &  10.35 &   -8.93 &   -8.65 &   -8.39 &  0.35 \\
UGC3432          &   6.0 &   8.32 &   9.22 &   9.55 &   9.17 &  10.15 &  10.45 &  -10.65 &   -9.76 &   -8.43 &  0.20 \\
UGC3576          &   3.0 &   8.81 &   9.14 &   9.77 &  10.09 &  10.57 &  11.43 &  -11.32 &  -10.16 &   -8.94 &  0.10 \\
UGC3725          &  -3.0 &   9.49 &   9.99 &  10.10 &  11.28 &  12.24 &  12.34 &  -12.00 &  -11.18 &   -9.80 &  0.50 \\
UGC3770          &  10.0 &   7.00 &   7.51 &   7.93 &   9.65 &   9.82 &  10.10 &   -9.11 &   -8.65 &   -8.18 &  0.40 \\
UGC3845          &   3.9 &   7.00 &   7.43 &   7.65 &   9.16 &   9.30 &   9.39 &   -8.89 &   -8.65 &   -8.41 &  0.20 \\
UGC4008          &   0.0 &   7.57 &   9.30 &   9.58 &  10.85 &  10.98 &  11.12 &  -10.80 &  -10.43 &   -8.54 &  0.10 \\
UGC4195          &   3.0 &   8.65 &   9.21 &   9.25 &  10.35 &  10.50 &  10.54 &   -9.97 &   -9.72 &   -9.23 &  0.25 \\
UGC4322          &  -5.0 &   8.07 &   8.31 &   9.89 &  10.64 &  11.12 &  12.02 &  -12.00 &   -8.61 &   -8.04 &  0.70 \\
UGC4414          &   0.0 &   9.09 &   9.46 &   9.56 &  10.86 &  10.95 &  11.06 &  -10.78 &  -10.64 &  -10.11 &  0.10 \\
UGC4455          &   1.0 &   8.66 &   9.22 &   9.67 &  10.61 &  10.76 &  10.90 &  -10.78 &   -9.96 &   -9.17 &  0.15 \\
UGC4614          & \ldots &   7.23 &   8.32 &   9.21 &   9.96 &  10.32 &  10.49 &   -9.94 &   -8.97 &   -8.31 &  0.30 \\
UGC5129          &   1.0 &   8.62 &   8.98 &   9.35 &  10.11 &  10.22 &  10.34 &  -10.59 &   -9.72 &   -9.09 &  0.15 \\
UGC52            &   5.0 &   8.19 &   9.32 &   9.49 &  10.24 &  10.38 &  10.56 &  -10.74 &  -10.45 &   -8.76 &  0.00 \\
UGC5234          &   5.0 &   7.51 &   8.90 &   9.42 &  10.15 &  10.55 &  10.80 &  -10.62 &   -9.44 &   -8.34 &  0.25 \\
UGC5378          &   3.0 &   7.20 &   8.28 &   9.21 &   9.52 &   9.88 &  10.09 &  -10.19 &   -8.94 &   -8.27 &  0.35 \\
UGC5542          &  -5.0 &   9.51 &   9.61 &  10.05 &  10.92 &  10.99 &  11.20 &  -12.00 &  -10.88 &  -10.68 &  0.05 \\
UGC6332          &   1.0 &   8.98 &   9.35 &   9.84 &  10.66 &  10.76 &  10.96 &  -12.00 &  -10.49 &   -9.64 &  0.10 \\
UGC7357          &   5.0 &   8.21 &   8.91 &   9.31 &   9.82 &   9.88 &  10.19 &  -10.01 &   -9.62 &   -8.71 &  0.00 \\
UGC7934          &   2.8 &   8.67 &   9.10 &   9.36 &  10.29 &  10.46 &  10.57 &  -10.33 &   -9.63 &   -9.19 &  0.20 \\
UGC8162          &   6.0 &   8.07 &   8.56 &   9.21 &   9.93 &  10.06 &  10.21 &   -9.75 &   -9.18 &   -8.78 &  0.15 \\
UGC9391          &   8.0 &   8.33 &   8.76 &   9.22 &   8.47 &   8.60 &   8.74 &   -9.78 &   -9.36 &   -8.96 &  0.05 \\
UGC9612          &   5.0 &   7.45 &   8.44 &   9.20 &  10.43 &  10.70 &  10.89 &  -10.12 &   -9.06 &   -8.43 &  0.30 \\
UGC9640          &  -5.0 &   9.51 &   9.70 &  10.05 &  11.39 &  11.48 &  11.65 &  -12.00 &  -12.00 &  -10.72 &  0.05
\tablenotetext{1}{Numerical type according to \citet{Vaucouleurs:59:275}}
\end{longtable}

\clearpage

\begin{longtable}{llrrrr}
\tablecaption{$^{56}$Ni Mass And Host Metallicity\label{tab_ni_metal}}
\tablewidth{0pt}
\tablehead{
 & & & \colhead{Corrected} \\
 & & \colhead{$^{56}$Ni Mass} & \colhead{$^{56}$Ni Mass} & \colhead{Mass-Metal} & \colhead{Prieto et al. 2006}\\
\colhead{SN} & \colhead{HOST} &
\colhead{(M$_{\odot}$)} & \colhead{(M$_{\odot}$)} &
\colhead{12 + $\log$(O/H)} & \colhead{12 + $\log$(O/H)}
}
1999gd   & NGC2623          &   0.15$\pm$0.02 &   0.34$\pm$0.04 &   9.10 &  \ldots \\
2006cc   & UGC10244         &   0.20$\pm$0.02 &   0.68$\pm$0.06 &   9.12 &  \ldots \\
2007ci   & NGC3873          &   0.26$\pm$0.02 &   0.31$\pm$0.03 &   9.19 &  \ldots \\
2001N    & NGC3327          &   0.27$\pm$0.03 &   0.57$\pm$0.06 &   9.16 &  \ldots \\
2005kc   & NGC7311          &   0.27$\pm$0.03 &   0.51$\pm$0.05 &   9.18 &  \ldots \\
2001da   & NGC7780          &   0.29$\pm$0.04 &   0.30$\pm$0.04 &   9.10 &  \ldots \\
2006bq   & NGC6685          &   0.30$\pm$0.03 &   0.41$\pm$0.04 &   9.16 &  \ldots \\
2006ar   & M+11-13-36       &   0.30$\pm$0.03 &   0.47$\pm$0.04 &   8.95 &   9.09 \\
2004L    & M+03-27-38       &   0.31$\pm$0.03 &   0.45$\pm$0.05 &   9.10 &  \ldots \\
2007bz   & IC3918           &   0.31$\pm$0.03 &   0.52$\pm$0.05 &   8.84 &  \ldots \\
2005ls   & M+07-07-01       &   0.34$\pm$0.03 &   0.79$\pm$0.08 &   8.99 &  \ldots \\
2008af   & UGC9640          &   0.35$\pm$0.05 &   0.51$\pm$0.07 &   9.21 &  \ldots \\
1996bo   & NGC673           &   0.36$\pm$0.03 &   0.75$\pm$0.07 &   9.10 &  \ldots \\
1994M    & NGC4493          &   0.36$\pm$0.04 &   0.35$\pm$0.04 &   9.19 &  \ldots \\
2006N    & M+11-08-12       &   0.36$\pm$0.04 &   0.35$\pm$0.03 &   9.17 &  \ldots \\
1999cc   & NGC6038          &   0.37$\pm$0.03 &   0.37$\pm$0.03 &   9.18 &  \ldots \\
2002he   & UGC4322          &   0.37$\pm$0.03 &   0.36$\pm$0.03 &   9.19 &  \ldots \\
2006al   & A103929+0511     &   0.38$\pm$0.06 &   0.42$\pm$0.07 &   9.08 &  \ldots \\
2003U    & NGC6365A         &   0.38$\pm$0.03 &   0.39$\pm$0.03 &   9.16 &  \ldots \\
2006ac   & NGC4619          &   0.38$\pm$0.03 &   0.56$\pm$0.05 &   9.18 &  \ldots \\
2007bc   & UGC6332          &   0.38$\pm$0.04 &   0.38$\pm$0.03 &   9.16 &  \ldots \\
2004as   & A112539+2249     &   0.39$\pm$0.04 &   0.44$\pm$0.04 &   8.81 &  \ldots \\
2001ie   & UGC5542          &   0.40$\pm$0.09 &   0.45$\pm$0.10 &   9.19 &  \ldots \\
2006ej   & NGC191A          &   0.41$\pm$0.04 &   0.47$\pm$0.05 &   9.19 &  \ldots \\
2006sr   & UGC14            &   0.41$\pm$0.04 &   0.42$\pm$0.04 &   9.15 &  \ldots \\
2006bt   & M+03-41-04       &   0.42$\pm$0.04 &   0.74$\pm$0.07 &   9.19 &  \ldots \\
2003W    & UGC5234          &   0.43$\pm$0.04 &   0.59$\pm$0.05 &   9.13 &  \ldots \\
2005ki   & NGC3332          &   0.44$\pm$0.04 &   0.37$\pm$0.03 &   9.20 &  \ldots \\
2002bf   & CGCG266-031      &   0.44$\pm$0.04 &   0.67$\pm$0.07 &   9.14 &  \ldots \\
2005iq   & M-03-01-08       &   0.45$\pm$0.04 &   0.40$\pm$0.04 &   9.10 &  \ldots \\
2007bd   & UGC4455          &   0.45$\pm$0.04 &   0.42$\pm$0.04 &   9.16 &  \ldots \\
1992bl   & E291-G11         &   0.45$\pm$0.05 &   0.40$\pm$0.04 &   9.20 &  \ldots \\
2002de   & NGC6104          &   0.45$\pm$0.04 &   0.67$\pm$0.07 &   9.17 &  \ldots \\
1992ag   & E508-G67         &   0.47$\pm$0.05 &   0.61$\pm$0.06 &   9.03 &  \ldots \\
2006cp   & UGC7357          &   0.47$\pm$0.04 &   0.59$\pm$0.05 &   8.99 &  \ldots \\
2006S    & UGC7934          &   0.48$\pm$0.04 &   0.69$\pm$0.06 &   9.12 &  \ldots \\
2006en   & M+05-54-41       &   0.50$\pm$0.06 &   0.66$\pm$0.09 &   9.15 &  \ldots \\
2004eo   & NGC6928          &   0.50$\pm$0.05 &   0.58$\pm$0.05 &   9.20 &  \ldots \\
2000fa   & UGC3770          &   0.51$\pm$0.05 &   0.58$\pm$0.05 &   8.98 &  \ldots \\
2007O    & UGC9612          &   0.51$\pm$0.05 &   0.54$\pm$0.06 &   9.15 &  \ldots \\
2002jy   & NGC477           &   0.51$\pm$0.05 &   0.52$\pm$0.05 &   9.12 &  \ldots \\
2006az   & NGC4172          &   0.51$\pm$0.04 &   0.47$\pm$0.04 &   9.20 &  \ldots \\
2002ha   & NGC6962          &   0.53$\pm$0.05 &   0.47$\pm$0.04 &   9.19 &   8.94 \\
2006on   & A215558-0104     &   0.53$\pm$0.09 &   0.67$\pm$0.11 &   9.09 &  \ldots \\
1996C    & M+08-25-47       &   0.53$\pm$0.05 &   0.66$\pm$0.06 &   9.02 &  \ldots \\
2001en   & NGC523           &   0.54$\pm$0.10 &   0.57$\pm$0.10 &   9.10 &  \ldots \\
2005ms   & UGC4614          &   0.54$\pm$0.05 &   0.55$\pm$0.05 &   9.09 &   9.06 \\
1996bv   & UGC3432          &   0.57$\pm$0.06 &   0.77$\pm$0.08 &   9.06 &  \ldots \\
2006te   & A081144+4133     &   0.57$\pm$0.07 &   0.50$\pm$0.06 &   9.09 &   9.10 \\
1999dk   & UGC1087          &   0.58$\pm$0.05 &   0.68$\pm$0.06 &   9.07 &  \ldots \\
2007F    & UGC8162          &   0.60$\pm$0.05 &   0.60$\pm$0.05 &   9.04 &   9.05 \\
1998dx   & UGC11149         &   0.61$\pm$0.06 &   0.50$\pm$0.05 &   9.20 &  \ldots \\
2006ax   & NGC3663          &   0.62$\pm$0.05 &   0.56$\pm$0.05 &   9.17 &  \ldots \\
2001fe   & UGC5129          &   0.64$\pm$0.06 &   0.70$\pm$0.06 &   9.07 &  \ldots \\
1992P    & IC3690           &   0.66$\pm$0.08 &   0.54$\pm$0.07 &   9.10 &  \ldots \\
1998ab   & NGC4704          &   0.67$\pm$0.06 &   0.72$\pm$0.06 &   9.14 &   9.16 \\
2001ba   & M-05-28-01       &   0.67$\pm$0.06 &   0.54$\pm$0.05 &   9.18 &  \ldots \\
1994S    & NGC4495          &   0.67$\pm$0.06 &   0.61$\pm$0.06 &   9.13 &  \ldots \\
1998V    & NGC6627          &   0.70$\pm$0.08 &   0.69$\pm$0.08 &   9.17 &  \ldots \\
1999aa   & NGC2595          &   0.71$\pm$0.06 &   0.64$\pm$0.05 &   9.16 &  \ldots \\
1991U    & IC4232           &   0.72$\pm$0.08 &   0.68$\pm$0.08 &   9.19 &  \ldots \\
2001az   & UGC10483         &   0.72$\pm$0.08 &   0.65$\pm$0.07 &   9.15 &  \ldots \\
2005eq   & M-01-09-06       &   0.72$\pm$0.06 &   0.98$\pm$0.09 &   9.14 &  \ldots \\
2007ae   & UGC10704         &   0.74$\pm$0.08 &   0.86$\pm$0.09 &   9.21 &  \ldots \\
2002ck   & UGC10030         &   0.74$\pm$0.10 &   0.71$\pm$0.10 &   9.18 &   9.23 \\
2008bf   & NGC4055          &   0.74$\pm$0.06 &   0.94$\pm$0.08 &   9.21 &  \ldots \\
2002hu   & M+06-06-12       &   0.80$\pm$0.07 &   0.73$\pm$0.06 &   9.08 &  \ldots \\
2000ca   & E383-G32         &   0.87$\pm$0.08 &   0.76$\pm$0.07 &   9.03 &  \ldots \\
2001V    & NGC3987          &   0.88$\pm$0.07 &   1.05$\pm$0.09 &   9.16 &  \ldots \\
1991ag   & IC4919           &   0.89$\pm$0.10 &   0.86$\pm$0.09 &   8.73 &  \ldots \\
1992bc   & E300-G09         &   0.90$\pm$0.08 &   0.71$\pm$0.06 &   8.95 &  \ldots \\
2003fa   & M+07-36-33       &   0.92$\pm$0.08 &   1.04$\pm$0.09 &   9.17 &  \ldots \\
2006cj   & A125924+2820     &   0.95$\pm$0.12 &   0.85$\pm$0.11 &   9.11 &  \ldots \\
1999dq   & NGC976           &   0.96$\pm$0.09 &   1.22$\pm$0.11 &   9.16 &  \ldots
\end{longtable}



\begin{thebibliography}{57}
\expandafter\ifx\csname natexlab\endcsname\relax\def\natexlab#1{#1}\fi

\bibitem[{Arnett(1979)}]{Arnett:79:L37}
Arnett, W.~D. 1979, \apj, 230, L37

\bibitem[{Arnett(1982)}]{Arnett:82:785}
---. 1982, \apj, 253, 785

\bibitem[{Astier {et~al.}(2006)Astier, Guy, Regnault, Pain, Aubourg, Balam,
  Basa, Carlberg, {et~al.}}]{Astier:06:31}
Astier, P., Guy, J., Regnault, N., Pain, R., Aubourg, E., Balam, D., Basa, S.,
  Carlberg, R.~G., {et~al.} 2006, \aap, 447, 31

\bibitem[{Bronder {et~al.}(2008)Bronder, Hook, Astier, Balam, Balland, Basa,
  Carlberg, Conley, {et~al.}}]{Bronder:08:717}
Bronder, T.~J., Hook, I.~M., Astier, P., Balam, D., Balland, C., Basa, S.,
  Carlberg, R.~G., Conley, A.~J., {et~al.} 2008, \aap, 477, 717

\bibitem[{Calzetti {et~al.}(1994)Calzetti, Kinney, \&
  Storchi-Bergmann}]{Calzetti:94:582}
Calzetti, D., Kinney, A.~L., \& Storchi-Bergmann, T. 1994, \apj, 429, 582

\bibitem[{Colgate \& McKee(1969)}]{Colgate:69:623}
Colgate, S.~A. \& McKee, C. 1969, \apj, 157, 623

\bibitem[{Conley {et~al.}(2007)Conley, Carlberg, Guy, Howell, Jha, Riess, \&
  Sullivan}]{Conley:07:L13}
Conley, A., Carlberg, R.~G., Guy, J., Howell, D.~A., Jha, S., Riess, A.~G., \&
  Sullivan, M. 2007, \apj, 664, L13

\bibitem[{Conley {et~al.}(2009)}]{Conley:09}
Conley, A. {et~al.} 2009, in preparation

\bibitem[{Conley {et~al.}(2008)Conley, Sullivan, Hsiao, Guy, Astier, Balam,
  Balland, Basa, {et~al.}}]{Conley:08:482}
Conley, A.~J., Sullivan, M., Hsiao, E.~Y., Guy, J., Astier, P., Balam, D.,
  Balland, C., Basa, S., {et~al.} 2008, \apj, 681, 482

\bibitem[{Fernandes {et~al.}(2005)Fernandes, Mateus, Sodr{\'e}, Stasi{\'n}ska,
  \& Gomes}]{Fernandes:05:363}
Fernandes, R.~C., Mateus, A., Sodr{\'e}, L., Stasi{\'n}ska, G., \& Gomes, J.~M.
  2005, \mnras, 358, 363

\bibitem[{Fioc \& Rocca-Volmerange(1997)}]{Fioc:97:950}
Fioc, M. \& Rocca-Volmerange, B. 1997, \aap, 326, 950

\bibitem[{Foley {et~al.}(2008)Foley, Matheson, Blondin, Chornock, Silverman,
  Challis, Clocchiatti, Filippenko, {et~al.}}]{Foley:08}
Foley, R.~J., Matheson, T., Blondin, S., Chornock, R., Silverman, J.~M.,
  Challis, P., Clocchiatti, A., Filippenko, A.~V., {et~al.} 2008,
  arXiv:astro-ph, astro-ph

\bibitem[{Gallagher {et~al.}(2005)Gallagher, Garnavich, Berlind, Challis, Jha,
  \& Kirshner}]{Gallagher:05:210}
Gallagher, J.~S., Garnavich, P.~M., Berlind, P., Challis, P., Jha, S., \&
  Kirshner, R.~P. 2005, \apj, 634, 210

\bibitem[{Gallagher {et~al.}(2008)Gallagher, Garnavich, Caldwell, Kirshner,
  Jha, Li, Ganeshalingam, \& Filippenko}]{Gallagher:08:752}
Gallagher, J.~S., Garnavich, P.~M., Caldwell, N., Kirshner, R.~P., Jha, S.~W.,
  Li, W., Ganeshalingam, M., \& Filippenko, A.~V. 2008, \apj, 685, 752

\bibitem[{Garnavich {et~al.}(2004)Garnavich, Bonanos, Krisciunas, Jha,
  Kirshner, Schlegel, Challis, Macri, {et~al.}}]{Garnavich:04:1120}
Garnavich, P.~M., Bonanos, A.~Z., Krisciunas, K., Jha, S., Kirshner, R.~P.,
  Schlegel, E.~M., Challis, P., Macri, L.~M., {et~al.} 2004, \apj, 613, 1120

\bibitem[{Gil~de~Paz {et~al.}(2007)Gil~de~Paz, Boissier, Madore, Seibert, Joe, Boselli,
  Wyder, Thilker, {et~al.}}]{Paz:07:185}
Gil~de~Paz, A., Boissier, S., Madore, B.~F., Seibert, M., Joe, Y.~H., Boselli,
  A., Wyder, T.~K., Thilker, D., {et~al.} 2007, \apjs, 173, 185

\bibitem[{Gonzalez-Gaitan {et~al.}(2009)}]{Gaitan:09}
Gonzalez-Gaitan {et~al.} 2009, in preparation

\bibitem[{Guy {et~al.}(2005)Guy, Astier, Nobili, Regnault, \&
  Pain}]{Guy:05:781}
Guy, J., Astier, P., Nobili, S., Regnault, N., \& Pain, R. 2005, \aap, 443, 781

\bibitem[{Hamuy {et~al.}(1995)Hamuy, Phillips, Maza, Suntzeff, Schommer, \&
  Aviles}]{Hamuy:95:1}
Hamuy, M., Phillips, M.~M., Maza, J., Suntzeff, N.~B., Schommer, R.~A., \&
  Aviles, R. 1995, \aj, 109, 1

\bibitem[{Hamuy {et~al.}(2000)Hamuy, Trager, Pinto, Phillips, Schommer, Ivanov,
  \& Suntzeff}]{Hamuy:00:1479}
Hamuy, M., Trager, S.~C., Pinto, P.~A., Phillips, M.~M., Schommer, R.~A.,
  Ivanov, V., \& Suntzeff, N.~B. 2000, \aj, 120, 1479

\bibitem[{Hicken {et~al.}(2009{\natexlab{a}})Hicken, Challis, Jha, Kirshner,
  Matheson, Modjaz, Rest, \& Wood-Vasey}]{Hicken:09b}
Hicken, M., Challis, P., Jha, S., Kirshner, R.~P., Matheson, T., Modjaz, M.,
  Rest, A., \& Wood-Vasey, W.~M. 2009{\natexlab{a}}, arXiv:astro-ph,
  astro-ph.CO

\bibitem[{Hicken {et~al.}(2009{\natexlab{b}})Hicken, Wood-Vasey, Blondin,
  Challis, Jha, Kelly, Rest, \& Kirshner}]{Hicken:09a}
Hicken, M., Wood-Vasey, W.~M., Blondin, S., Challis, P., Jha, S., Kelly, P.~L.,
  Rest, A., \& Kirshner, R.~P. 2009{\natexlab{b}}, arXiv:astro-ph, astro-ph.CO

\bibitem[{Howell(2001)}]{Howell:01:L193}
Howell, D.~A. 2001, \apj, 554, L193

\bibitem[{Howell {et~al.}(2009)Howell, Sullivan, Brown, Conley, Le~Borgne,
  Hsiao, Astier, Balam, {et~al.}}]{Howell:09:661}
Howell, D.~A., Sullivan, M., Brown, E.~F., Conley, A., Le~Borgne, D., Hsiao,
  E.~Y., Astier, P., Balam, D., {et~al.} 2009, \apj, 691, 661

\bibitem[{Howell {et~al.}(2007)Howell, Sullivan, Conley, \&
  Carlberg}]{Howell:07:L37}
Howell, D.~A., Sullivan, M., Conley, A., \& Carlberg, R. 2007, \apj, 667, L37

\bibitem[{Jha {et~al.}(2007)Jha, Riess, \& Kirshner}]{Jha:07:122}
Jha, S., Riess, A.~G., \& Kirshner, R.~P. 2007, \apj, 659, 122

\bibitem[{Kasen(2006)}]{Kasen:06:939}
Kasen, D. 2006, \apj, 649, 939

\bibitem[{Kobayashi \& Nomoto(2007)}]{Kobayashi:07:215}
Kobayashi, C. \& Nomoto, K. 2007, arXiv:astro-ph, 0801, 215

\bibitem[{Le~Borgne \& Rocca-Volmerange(2002)}]{Borgne:02:446}
Le~Borgne, D. \& Rocca-Volmerange, B. 2002, \aap, 386, 446

\bibitem[{Le~Borgne {et~al.}(2004)Le~Borgne, Rocca-Volmerange, Prugniel, Lan{\c c}on,
  Fioc, \& Soubiran}]{Borgne:04:881}
Le~Borgne, D., Rocca-Volmerange, B., Prugniel, P., Lan{\c c}on, A., Fioc, M.,
  \& Soubiran, C. 2004, \aap, 425, 881

\bibitem[{Lee {et~al.}(2006)Lee, Skillman, Cannon, Jackson, Gehrz, Polomski, \&
  Woodward}]{Lee:06:970}
Lee, H., Skillman, E.~D., Cannon, J.~M., Jackson, D.~C., Gehrz, R.~D.,
  Polomski, E.~F., \& Woodward, C.~E. 2006, \apj, 647, 970

\bibitem[{Martin {et~al.}(2005)Martin, Fanson, Schiminovich, Morrissey,
  Friedman, Barlow, Conrow, Grange, {et~al.}}]{Martin:05:L1}
Martin, D.~C., Fanson, J., Schiminovich, D., Morrissey, P., Friedman, P.~G.,
  Barlow, T.~A., Conrow, T., Grange, R., {et~al.} 2005, \apj, 619, L1

\bibitem[{Parodi {et~al.}(2000)Parodi, Saha, Sandage, \&
  Tammann}]{Parodi:00:634}
Parodi, B.~R., Saha, A., Sandage, A., \& Tammann, G.~A. 2000, \apj, 540, 634

\bibitem[{Perlmutter {et~al.}(1999)Perlmutter, Aldering, Goldhaber, Knop,
  Nugent, Castro, Deustua, Fabbro, {et~al.}}]{Perlmutter:99:565}
Perlmutter, S., Aldering, G., Goldhaber, G., Knop, R.~A., Nugent, P., Castro,
  P.~G., Deustua, S., Fabbro, S., {et~al.} 1999, \apj, 517, 565

\bibitem[{Perlmutter {et~al.}(1997)Perlmutter, Gabi, Goldhaber, Goobar, Groom,
  Hook, Kim, Kim, {et~al.}}]{Perlmutter:97:565}
Perlmutter, S., Gabi, S., Goldhaber, G., Goobar, A., Groom, D.~E., Hook, I.~M.,
  Kim, A.~G., Kim, M.~Y., {et~al.} 1997, \apj, 483, 565

\bibitem[{Phillips(1993)}]{Phillips:93:L105}
Phillips, M.~M. 1993, \apj, 413, L105

\bibitem[{Prieto {et~al.}(2006)Prieto, Rest, \& Suntzeff}]{Prieto:06:501}
Prieto, J.~L., Rest, A., \& Suntzeff, N.~B. 2006, \apj, 647, 501

\bibitem[{Prieto {et~al.}(2008)Prieto, Stanek, \& Beacom}]{Prieto:08:999}
Prieto, J.~L., Stanek, K.~Z., \& Beacom, J.~F. 2008, \apj, 673, 999

\bibitem[{Riess {et~al.}(1998)Riess, Filippenko, Challis, Clocchiatti, Diercks,
  Garnavich, Gilliland, Hogan, {et~al.}}]{Riess:98:1009}
Riess, A.~G., Filippenko, A.~V., Challis, P., Clocchiatti, A., Diercks, A.,
  Garnavich, P.~M., Gilliland, R.~L., Hogan, C.~J., {et~al.} 1998, \aj, 116,
  1009

\bibitem[{Riess {et~al.}(1996)Riess, Press, \& Kirshner}]{Riess:96:88}
Riess, A.~G., Press, W.~H., \& Kirshner, R.~P. 1996, \apj, 473, 88

\bibitem[{Schawinski(2009)}]{Schawinski:09}
Schawinski, K. 2009, arXiv:astro-ph, astro-ph.CO

\bibitem[{Schlegel {et~al.}(1998)Schlegel, Finkbeiner, \&
  Davis}]{Schlegel:98:525}
Schlegel, D.~J., Finkbeiner, D.~P., \& Davis, M. 1998, \apj, 500, 525

\bibitem[{Seibert {et~al.}(2009)}]{Seibert:09}
Seibert, M. {et~al.} 2009, in preparation

\bibitem[{Sullivan(2009)}]{Sullivan:09:539}
Sullivan, M. 2009, Probing Stellar Populations Out To The Distant Universe:
  CEFALU 2008, 1111, 539

\bibitem[{Sullivan {et~al.}(2006)Sullivan, Le~Borgne, Pritchet, Hodsman, Neill,
  Howell, Carlberg, Astier, {et~al.}}]{Sullivan:06:868}
Sullivan, M., Le~Borgne, D., Pritchet, C.~J., Hodsman, A., Neill, J.~D.,
  Howell, D.~A., Carlberg, R.~G., Astier, P., {et~al.} 2006, \apj, 648, 868

\bibitem[{Sullivan {et~al.}(2009)Sullivan, Ellis, Howell, Riess, Nugent, \&
  Gal-Yam}]{Sullivan:09b}
Sullivan, M., Ellis, R.~S., Howell, D.~A., Riess, A., Nugent, P.~E., \&
  Gal-Yam, A. 2009, arXiv:astro-ph, astro-ph.SR

\bibitem[{Taubenberger {et~al.}(2008)Taubenberger, Hachinger, Pignata, Mazzali,
  Contreras, Valenti, Pastorello, Elias-Rosa, {et~al.}}]{Taubenberger:08:75}
Taubenberger, S., Hachinger, S., Pignata, G., Mazzali, P.~A., Contreras, C.,
  Valenti, S., Pastorello, A., Elias-Rosa, N., {et~al.} 2008, \mnras, 385, 75

\bibitem[{Timmes {et~al.}(2003)Timmes, Brown, \& Truran}]{Timmes:03:L83}
Timmes, F.~X., Brown, E.~F., \& Truran, J.~W. 2003, \apj, 590, L83

\bibitem[{Tonry {et~al.}(2003)Tonry, Schmidt, Barris, Candia, Challis,
  Clocchiatti, Coil, Filippenko, {et~al.}}]{Tonry:03:1}
Tonry, J.~L., Schmidt, B.~P., Barris, B., Candia, P., Challis, P., Clocchiatti,
  A., Coil, A.~L., Filippenko, A.~V., {et~al.} 2003, \apj, 594, 1

\bibitem[{Tremonti {et~al.}(2004)Tremonti, Heckman, Kauffmann, Brinchmann,
  Charlot, White, Seibert, Peng, {et~al.}}]{Tremonti:04:898}
Tremonti, C.~A., Heckman, T.~M., Kauffmann, G., Brinchmann, J., Charlot, S.,
  White, S. D.~M., Seibert, M., Peng, E.~W., {et~al.} 2004, \apj, 613, 898

\bibitem[{Tripp(1998)}]{Tripp:98:815}
Tripp, R. 1998, \aap, 331, 815

\bibitem[{Tripp \& Branch(1999)}]{Tripp:99:209}
Tripp, R. \& Branch, D. 1999, \apj, 525, 209

\bibitem[{Truran {et~al.}(1967)Truran, Arnett, \& Cameron}]{Truran:67:2315}
Truran, J.~W., Arnett, W.~D., \& Cameron, A. G.~W. 1967, CJP, 45, 2315

\bibitem[{de~Vaucouleurs(1959)}]{Vaucouleurs:59:275}
de~Vaucouleurs, G. 1959, Handbuch der Physik, 53, 275

\bibitem[{de~Vaucouleurs {et~al.}(1991)de~Vaucouleurs, de~Vaucouleurs, Corwin,
  Buta, Paturel, \& Fouque}]{Vaucouleurs:91}
de~Vaucouleurs, G., de~Vaucouleurs, A., Corwin, H.~G., Buta, R.~J., Paturel,
  G., \& Fouque, P. 1991, Volume 1-3

\bibitem[{Wood-Vasey {et~al.}(2008)Wood-Vasey, Friedman, Bloom, Hicken, Modjaz,
  Kirshner, Starr, Blake, {et~al.}}]{Vasey:08:377}
Wood-Vasey, W.~M., Friedman, A.~S., Bloom, J.~S., Hicken, M., Modjaz, M.,
  Kirshner, R.~P., Starr, D.~L., Blake, C.~H., {et~al.} 2008, \apj, 689, 377

\bibitem[{York {et~al.}(2000)York, Adelman, Anderson, Anderson, Annis, Bahcall,
  Bakken, Barkhouser, {et~al.}}]{York:00:1579}
York, D.~G., Adelman, J., Anderson, J.~E., Anderson, S.~F., Annis, J., Bahcall,
  N.~A., Bakken, J.~A., Barkhouser, R., {et~al.} 2000, \aj, 120, 1579

\end{thebibliography}
\end{document}